\documentstyle[11pt,aaspp4]{article}  % preprint format

\begin{document}

\title{Contact Binaries of the Galactic Disk: \\
Comparison of the Baade's Window and Open Cluster Samples}

\author{\sc Slavek M. Rucinski\\
\rm Electronic-mail: {\it rucinski@cfht.hawaii.edu\/}}

\affil{Canada -- France -- Hawaii Telescope Co.\\
P.O.\ Box 1597, Kamuela, HI 96743}
 
\bigskip
\centerline{\today}
 
\bigskip
\begin{abstract}
The paper attempts to integrate the available data
for contact binaries of the disk population in a deep galactic field
and in old open clusters. The two basic data sets consist
of 98 systems in the volume-limited 3 kpc sub-sample
of contact binaries detected by the OGLE microlensing project
toward Baade's Window (BW$_3$) and of 63 members of 
11 old open clusters (CL). Supplementary
data on the intrinsically bright, but spatially rare, long-period
binaries are provided by 238 systems in the BW sample  
to the distance of 5 kpc (BW$_5$). The basic BW$_3$ sample and 
the CL sample are remarkably similar in the period, color, 
luminosity and variability-amplitude distributions, in spite
of very different selections, for BW$_3$ -- 
as a volume-limited sub-sample
of all contact systems discovered by the OGLE project,
and for CL -- as a collection of contact systems discovered
in open clusters which had been subject to searches differing
in limiting magnitudes, cluster area coverage and photometric errors.
The contact systems are found in the color interval
$0.3 < (B-V)_0 < 1.2$ where the turn-off points (TOP) of the
considered clusters are located; however, they are not 
concentrated at the respective TOP locations but, once 
the TOP happens to fall in the above color interval, they
can appear anywhere within it. 

The luminosity function for the BW sample appears to be 
very similar in shape to that for the solar neighborhood 
main-sequence (MS) stars when corrections for the galactic
disk structure are applied, implying a flat apparent 
frequency-of-occurrence distribution. In the accessible 
interval $2.5 < M_V < 7.5$, the frequency of contact binaries
relative to MS stars equals about 1/130 for the
exponential disk length scale $h_R = 2.5$ kpc and   
about 1/100 for $h_R = 3.5$ kpc. 
The high frequency cannot continue for $M_V < 2.5$ as the predicted
numbers of bright systems would then become 
inconsistent with the numbers of known systems
to $V_{lim} = 7.5$ in the sky sample. The 
previous estimate of the frequency from the
BW sample of $1/250 - 1/300$ did not correctly relate 
the numbers of the contact binaries 
to the numbers of MS stars. The magnitude 
limit of the OGLE survey limits the accuracy of the current
luminosity function determination for $M_V > 5.5$, 
but the available data are consistent with a continuation
of the high apparent frequency beyond $M_V = 7.5$, 
i.e.\ past the current short-period, low-luminosity end,
delineated by the shortest-period field system CC~Com at $M_V = 6.7$. 
The current data indicate that the sky-field sample starts showing 
discovery-selection effects at a level as high as $V \simeq 10 - 11$. 

\end{abstract}

\keywords{binaries: general -- binaries: eclipsing -- stars: statistics}
 
\section{INTRODUCTION}
\label{intro}

Contact stars are close binary systems in which components
form single entities described by equipotentials of 
the Roche geometry. The most common among them 
consist of solar-type stars and are called 
W~UMa-type binaries; their orbital periods 
are in the range between about one quarter and three quarters 
of a day. Several 
reviews have discussed properties of contact binaries; the recent 
ones, concentrating respectively on theoretical and observational
issues have been by Eggleton \markcite{egg96} (1996)
and by Rucinski \markcite{ruc93} (1993).
How these binaries form and evolve is still poorly 
understood, but it is generally assumed that they are in the 
penultimate -- but possibly long lasting -- stage of  
angular-momentum-loss driven evolution, 
just before forming single stars. The angular-momentum-loss 
results from the torque exerted by the magnetized stellar wind 
on the components, which extracts the 
angular momentum from the orbit via the tidal synchronism of 
rotation. Since this process takes relatively long time, 
of the order of a few Gyrs for solar-type stars, 
the W~UMa-type systems are expected to consist
of relatively old stars. The evidence for their advanced 
age cannot be 
inferred from spectral signatures of low metal abundance because of 
the extremely strong broadening of spectral lines, but comes 
from (1)~their relatively large spatial velocities 
(Guinan \& Bradstreet \markcite{gb88} 1988), characteristic for
old disk stars, and their presence in (2)~old open clusters 
(Ka\l u\.zny \& Rucinski \markcite{kr93a} 1993 = KR93, 
Rucinski \& Ka\l u\.zny \markcite{rk94} 1994 = RK94) and in
(3)~the disk component toward
Baade's Window (Rucinski \markcite{ruc97a} 1997a, see below).

For a long time, the statistics of contact binaries was particularly 
uncertain because of the accidental nature of 
the sky-field discoveries. One of the indications of
incompleteness in the cataloged sky-field sample was the tendency to show
only relatively large 
light-curve amplitudes whereas simple considerations of randomly 
distributed orbital inclinations suggest that low amplitude systems should
be most common. Indeed, systematic searches in open clusters 
(KR93, RK94), later supplemented by the OGLE microlensing
by-product data (Rucinski \markcite{ruc97a} 1997a = R97a, 
Rucinski \markcite{ruc97b} 1997b = R97b; see below), 
led to the discovery of many low amplitude systems. The cluster 
searches also permitted to address the 
question of the ages. A progression in numbers of such 
systems with the cluster age, in the sense of more systems in older 
clusters, supported the view that the contact systems form over 
time from close, detached binary systems. The initial
interpretations of the old open cluster data (KR93, RK94)
indicated the apparent\footnote{Unless specifically noted, 
the frequency of occurrence of contact systems
discussed in this paper is {\it apparent\/}, i.e.\ 
it is not corrected for missed systems with low orbital
inclination angles. It is expected that the correction factor is
of the order of 1.5 to 2.0, but it cannot be predicted a priori
as its value crucially depends on the mass-ratio distribution, 
which is unknown, see R97b. In the same sense, when we 
discuss ``complete'' samples later on, we mean samples of those systems
which are discoverable for a given minimum amplitude threshold.} 
frequency in clusters by an order of magnitude higher
than had been estimated for the sky field (Duerbeck
\markcite{due84} 1984), 
reaching perhaps one such a system per hundred main 
sequence stars. When projected non-members were removed from 
clusters at low galactic latitudes 
and data for several clusters were averaged to improve
the number statistics, the apparent frequency relative to the number
of monitored MS stars of spectral types F to K 
was estimated to be about one such a system per
250 -- 300 ordinary stars (Rucinski \markcite{ruc94} 1994 = CAL1).
Although these numbers were approximate,
they indicated that in a typical old open cluster, where some
thousand or so stars could be monitored for variability,
only a few contact systems would be normally found, 
a circumstance making any meaningful statistical inferences
difficult. Thus, the cluster data could not offer sound
statistics and larger, more uniform data sets were needed.

Very recently, massive discoveries of contact systems in 
microlensing surveys offered an abundant source of statistical 
data. Two thirds among 933 eclipsing binaries discovered in the 
OGLE-1 survey (R97a, R97b) have been classified as contact 
binaries. While the discoveries yielded unprecedented material 
for studies of contact binaries in various ways (R97a, R97b, 
Rucinski \markcite{ruc98} 1998 = R98), one aspect has been 
of particular importance in the present context: 
The OGLE data permitted to define a volume-limited sample 
of contact binaries, leading to the first unbiased estimates
of frequency distributions for such parameters as orbital periods,
colors or luminosities. The contact systems were found to 
appear in unexpectedly narrow ranges of periods, mainly
within $0.25 < P < 0.7$ day, and colors, $ 0.2 < (V-I)_0 < 1.5$
(several observational effects could make the observed 
color range slightly broader than in reality, 
see Section~\ref{col}). The range in colors -- coinciding
with location on the color--magnitude diagrams where solar-type,
old-disk stars start showing evolutionary effects --
led to a suggestion (R97a) that the properties of these
systems have an evolutionary relation to 
the Turn-Off Point (TOP) stage of evolution, when the 
components expand and enter into physical contact. This suggestion
is testable since the cluster data of the CL sample
permit to relate positions
of contact systems and TOP's in color--magnitude diagrams.

This paper discusses properties of contact 
binaries toward Baade's Window (BW) and in old
open clusters (CL), to uncover similarities and differences
in the two sets of data. The two samples are 
complementary, as the BW sample contains objects observed in a
uniform fashion, but with absolute magnitudes and implied 
distances determined through one particular calibration, 
while the CL sample consist of objects with independent
information on age, metallicity and distance, but from an
group of clusters which were selected in a somewhat
non-rigorous fashion. This paper utilizes also 
a new, very useful tool which had not been available before, 
the new $M_V = M_V(\log P, B-V)$ calibration which is
based on the Hipparcos parallaxes (Rucinski \& 
Duerbeck \markcite{rd97} 1997 = CAL5). The calibration 
obviates any needs to resort to absolute magnitude 
calibrations based on open clusters, whose use could
imply an obvious risk of a circular reasoning. We stress
that this paper utilizes the Hipparcos-based 
{\it calibration\/}, but does not utilize the Hipparcos 
{\it sample\/} of the contact binaries in the solar vicinity. 
The reason is the biased character of the
Hipparcos sample which, being magnitude-limited, consists primarily of 
intrinsically luminous systems. We note, that the 
Hipparcos observed all stars of the sky only to 
slightly beyond $V \simeq 7$ and only six contact binaries 
actually exist to this magnitude limit.
 
The paper is organized as follows: Section~\ref{samples} describes
the BW and CL samples. The next sections compare various 
properties and statistics for the two samples: 
the period (Section~\ref{per}) and color
(Section~\ref{col}) distributions,
the color -- magnitude diagrams (Section~\ref{cmd}), the period -- color
relations (Section~\ref{PC}) and the variability amplitudes 
(Section~\ref{ampl}). The important matter of the frequency of 
occurrence of the contact systems is addressed in Section~\ref{freq}, after a 
discussion of the related issue of the 
luminosity functions in Section~\ref{LF}.
The last  Section~\ref{concl} contains conclusions of the paper.

\section{THE TWO SAMPLES}
\label{samples}

\subsection{The Baade's Window sample (BW)}

The volume-limited BW sample has been defined in R97a. 
It consists of 98 systems with orbital periods
shorter than one day and distances smaller than 3 kpc, which
passed the Fourier light-curve shape filter.  
Comparison of the number densities for the volumes defined
by the distances of 2 kpc and 3 kpc indicated that the 3 kpc
sample is complete to its limiting absolute magnitude of $M_I =
4.5$, which -- for typical values of reddening and intrinsic colors --
translates into an approximate limit of $M_V \simeq 5.5$.
Among the BW-sample systems, most are genuine W~UMa-type systems
with approximately equally-deep eclipses, indicating
perfect thermal contact between components. 
However, two systems did pass the light-curve shape filter used to
select contact systems, yet had 
eclipse depth differences large enough to suspect poor 
thermal contact, or more likely, very close semi-detached 
configurations; the accompanying asymmetries of the maxima
suggest in such cases an on-going mass transfer (R97b). 
Thus two, among 98 systems, i.e.\ 
only about 2 percent of all contact systems appear to be of 
this type, although -- by being intrinsically more luminous than typical
W~UMa-type systems due to longer periods  --
they are much more common in the sky-field (or any other 
magnitude-limited) sample. For simplicity, we will call them Poor 
Thermal Contact (PTC) systems, remembering that these 
could be either contact systems with inhibited energy transfer or
very close semi-detached binaries.

The intrinsically bright, long-period systems can be observed
deeper in space. By selecting a deeper sample, we can analyze
the long-period and high-luminosity ends of the respective
distributions, sacrificing the statistics at the faint,
short period end. For that purpose, a second BW sample to 5 kpc
has been considered in this paper. It is based
on a 4.6 times larger volume than the 3 kpc
sample and contains 238 system, 8 of them
showing the PTC light curves. Its statistical properties are sound only for
systems with periods longer than about 0.55 day 
as the short-period ones are eliminated by their low
absolute magnitudes (see Figure~13 in R97a). The expected
completeness limit of the 5 kpc BW sample
is $M_V \simeq 4.2$ (thus the $3.5 < M_V < 4.5$ bin 
is partly affected in various statistics presented later). 
Whenever relevant, we will distinguish the two Baade's Window
sub-samples by subscripts,  BW$_3$ and BW$_5$, but normally,
for most considerations, the basic sample BW$_3$ will be used.

Only one system with a period longer than one day,
BW5.009 with $P=1.59$ day, from the sample discussed in
R98 could be formally
included in the BW$_5$ sample (its distance is about
4.3 kpc). However, its luminosity and distance 
are poorly known due to lack of the luminosity calibration
for very blue contact systems (the observed $V-I=1.01$ and 
the estimated $E_{V-I}=0.84$). This system is disregarded in 
the current paper.

The original data in the OGLE catalog consist of the maximum
brightness $I$ and $V-I$ magnitudes and colors, orbital periods $P$, 
amplitudes in $I$ and coordinate positions. 
The analysis presented in R97a and R97b added to these data the 
light-curve-decomposition Fourier coefficients as well as 
the distances, absolute magnitudes $M_I$ and reddening corrections 
$E_{V-I}$. These data for all contact binaries in the OGLE survey
are available in the form of extensive tables via Internet
at: http://www.cfht.hawaii.edu/$\sim$rucinski/rucinski.html
(the major tables of this paper are also in this location).
Since our basic 3 kpc volume-limited sample has not 
yet been published, but may have a more general
use, we present it in Table~\ref{tab1}. The table contains
the original data from OGLE as well as the
derived quantities in the $V$, $B-V$  
photometric system. The transformation of the photometric data,
rather than the use of the original $I$, $V-I$ data (which would
have many advantages because of usually higher accuracy and weaker 
sensitivity of that color to entirely unknown metallicities), 
has been mandated by the fact that most open clusters were observed
in the $V$, $B-V$ system, and that the Hipparcos calibration 
is available only in this system. 
The transformations were: $M_V = M_I + (V-I) - E_{V-I}$ 
for the absolute magnitudes; the reddening-corrected colors
$(V-I)_0$ have been transformed into $(B-V)_0$
with the main-sequence relations
of Bessell \markcite{bes1} \markcite{bes2} (1979, 1990).

\placetable{tab1}

\subsection{The cluster sample (CL)}

The open cluster sample (CL) has been obtained by combining 
the data published in several papers, most of them by Ka\l u\.zny 
and collaborators. The assumed cluster properties are 
listed in Table~\ref{tab2} which is arranged in the age progression, 
from the oldest to the youngest clusters. To obtain the best 
uniformity of the material, the values of reddening 
corrections $E_{B-V}$ and apparent distance moduli, $(m-M)_V$, 
were taken from the recent tabulation of Twarog et al. 
\markcite{twa97} (1997). 
The ages have been taken mostly from the original publications, 
and then adjusted slightly for consistency of the 
color-magnitude diagrams (Section~\ref{cmd}). The ages are only 
approximate and used here mainly to arrange the 
clusters into an age progression. References to the sources of the 
photometric data are given in the last column of Table~\ref{tab2}.

\placetable{tab2}

\placetable{tab3}

Table~\ref{tab3} lists the data for
the contact systems detected in the cluster 
fields. The systems are ordered by the new variable-star
designations, following the 71st, 72nd and 73rd 
Variable-Star Name lists (Kazarovets et al.\ \markcite{kaz93} 1993,
Kazarovets \& Samus \markcite{kaz95} \markcite{kaz97} 1995, 1997).
Since these designations are used for the first time
for many of the listed systems, 
the original names in the discovery papers are also given.
Most of the photometric data are available in the $V$, $B-V$ 
system. For those clusters which were observed in the $V-I$ 
system and for those with both colors available, 
the main-sequence color--color transformations 
to $B-V$ were used, as the $V-I$ data 
were usually of better quality than those in $B-V$. 

The Fourier light-curve shape filter was not applied to the CL
systems to verify their W~UMa-type characteristics. The main reason
was the partial phase coverage for many systems which frequently
resulted in erroneous values of the Fourier coefficients
and rejection of otherwise apparently genuine cluster members. 
For this reason,
the CL sample may contain a small admixture of other short-period
variable stars; in this sense, the internal consistency of the CL
sample is poorer than that of the BW sample.
The cluster membership was verified using the absolute--magnitude
Hipparcos calibration CAL5: 
$M_V^{cal} = -4.44\, \log P + 3.02\, (B-V)_0 + 0.12$,
where $(B-V)_0=(B-V)-E_{B-V}$. 
The absolute magnitudes listed in Table~\ref{tab3} have 
been obtained using the observed maximum-light magnitudes, 
$V_{max}$, and the cluster distance moduli, $(m-M)_V$:
$M_V = V_{max} - (m-M)_V$. Large deviations
$\Delta M_V = M_V - M_V^{cal}$ permitted to identify systems 
located in front or behind the respective clusters. The deviations
form a distribution which shows long tails of non-members on both
sides of the maximum, but which within $-2 < \Delta M_V < +2$
 can be rather well described by a Gaussian with the dispersion 
$\sigma = 0.47$ and with the mean at $-0.04$. The small shift
in the mean value
is gratifying as it shows that the Hipparcos and cluster samples
are mutually consistent. However, the dispersion of $\Delta M_V$ is large
when compared with that for the Hipparcos sample which showed an intrinsic 
scatter of $\sigma_{HIP} = 0.22$. We suspect that a large fraction of
the scatter comes from the uncertainties in the cluster
data. This is indicated by systematically smaller deviations for
some better observed clusters such as M~67. But notice also that some
apparently genuine cluster members (such as ER~Cep in NGC~188) 
in well-observed clusters do show large deviations. For some clusters,
the deviations may have been increased because
no account was made for differing, but usually poorly known
metallicities (as in the case
of Tom~2). Instead of inventing a system of weights for
individual clusters, it was decided simply
to widen the range for the membership
acceptance in $\Delta M_V$ to $\pm 1.0$, that is
to $\pm 2.1 \sigma$ of the $\Delta M_V$ distribution. While this way
some non-members may have entered to spoil our statistics, we note that
the application of the Hipparcos calibration resulted 
-- in most cases -- in smaller deviations in $M_V$ than in the
discovery papers where individual membership criteria  
were first discussed. Thus, many 
systems which would not pass the $\pm 1$ 
magnitude deviation filter on the basis of the older calibrations 
CAL1 and CAL2 (Rucinski \markcite{ruc94}
\markcite{ruc95} 1994, 1995)
can now be considered as cluster members.

In two cases, V514~Lyr (NGC~6791--V8) and IK~CMa (Be~33--V2),
the deviations are slightly larger than the adopted
threshold; in both cases $\Delta M_V = -1.05$. An
inconsistency has been committed here
by removing V514~Lyr from the CL sample, but
retaining IK~CMa. The basic photometric data for NGC~6791 
are sufficiently well known to verify that the membership 
of V514~Lyr to the cluster is quite unlikely.
In contrast, the large and poorly known 
reddening ($E_{B-V} \simeq 0.7$) 
and, possibly, the low metallicity of Be~33 leave a large margin 
of uncertainty in the cluster properties to retain IK~CMa as
a probable member. This is the only system in Be~33 which 
can be considered as a member (the other one, II~CMa =
Be~33--V1 is definitely not). Be~33 is the
youngest of the clusters in the sample, so that it is easy to
keep IK~CMa apart and see if it deviates in any other sense.
As far as we can see, the system belongs to this cluster.

The cluster surveys for variability 
have different depth and reach different
levels of absolute magnitudes. An attempt has been made to
estimate roughly the levels at which detection of variability
would become impossible because of the rapid increase of
errors for fainter stars. These limiting levels, $M_V^{lim}$,
were estimated on the basis 
the cluster distance moduli and the photometric-error data
given in most of the papers by Ka\l u\.zny
and collaborators as the points where the errors reached about 
0.05 mag.; for such errors, variables with amplitudes of about 0.1
mag should be still detectable.
The limiting $M_V^{lim}$ are given in Table~\ref{tab2} and shown later
in the color -- magnitude diagrams shown in Section~\ref{cmd}. Typically, 
the nominal depths are $M_V^{lim} \simeq 6 - 7$, 
with the exception of the distant 
and photometrically difficult clusters Tom~2 and Be~33, where the
limits are at the level of about $M_V \simeq 5$. It should be
stressed that these limits are approximate and somewhat subjective
so that the CL sample is much less rigorously defined, especially
at its faint end, than the BW sample. 

In forming the CL sample, no account has been made of the fact that 
some searches of open clusters gave no discoveries of contact systems. 
The number of failed searches is not known. 
Only one case has been published of 
such a failed search in 6 clusters (NGC 2360, 2420, 2506,
6802, 6819 and Mel~66) by Ka\l u\.zny \& Shara 
\markcite{ks88} (1988); in one of these clusters, NGC~6802, a very low
amplitude W~UMa system was subsequently found (Vidal \& Belmonte
\markcite{vid93} 1993). We have no explanation for the 
lack of contact systems in some clusters and we do not know
if this is a real phenomenon or some statistical or observational 
effect. Whatever is the cause, it clearly shows the limitations
of the CL sample which is less rigorously defined than the BW$_3$ 
sample.

\section{PERIOD DISTRIBUTION}
\label{per}

A comparison of the BW and CL samples is shown
in Figure~\ref{fig1}. Since we have 98 objects of the
BW$_3$ sample and 63 objects of the CL sample, this figure and
the following similar ones have the left and right side vertical scales 
scaled in proportion of 3:2 to take into
account approximately the difference in sizes of the two 
samples. The histograms in Figure~\ref{fig1} and in the following similar 
figures are not normalized in order to 
show the numbers of the W~UMa-type systems in each bin and
thus permit a direct judgment on the Poisson uncertainties involved. 
We can see that whatever statistical
properties we would like to analyze in this paper, the results will be
relevant to the most common contact binaries only; 
objects appearing at frequencies lower than a few
percent of the totality are expected to be missed.
The BW$_5$ sample consisting of 238 objects
has been added to improve the statistics for
the intrinsically rare, but bright, long-period systems which can
be seen to large distances. 

\placefigure{fig1}

Taking into account the large per-bin uncertainties in the histograms, 
the period distributions for the BW$_3$ and CL samples 
(Figure~\ref{fig1}) 
are surprisingly similar, especially when plotted in linear 
units of the orbital period (the left panel in the figure).
Both samples show sharp cutoffs at the orbital 
periods of about 0.22 -- 0.25 day, and 
maxima at about 0.35 -- 0.4 day. Systems with periods longer
than about 0.7 day are absent in BW$_3$ but do appear in BW$_5$
at the level of about 0.5 percent of all systems.
The short period cutoffs are located at 0.228 day, 
defined by BW4.040\footnote{We use the same convention as in
R97a in that the first digit gives the OGLE field, and the
number of the variable in the field is given after the period.},
and 0.225 day, defined by V702~Mon in Be~39; both are very close
to the current record of 0.221 day for the field system
CC~Com (Rucinski \markcite{ruc77} 1977).
Application of the two-distribution 
Kolmogorov -- Smirnov test gives only 0.6 percent
significance to the hypothesis that the BW and CL
distributions are different, when binned in linear units of the 
orbital period. The same significance with the logarithmic
binning is 3.5 percent. 

The period distribution for the BW$_5$ sample indicates 
that contact systems with periods longer than 0.7 day are
very rare, but not absent, as could be perhaps 
erroneously inferred from the BW$_3$ sample. They become
detectable when sufficiently large volume is searched.
While Figure~\ref{fig1} shows the numbers 
for BW$_5$ simply scaled by 5 relative to BW$_3$, to 
allow for the difference in the search volumes in an approximate
way, Figure~\ref{fig2} and
Table~\ref{tab4} give the exact relation in the form of
what we call the {\it period function\/}, PF.
It is an analogue of the luminosity function and gives the number
of contact binaries in constant intervals of $\log P$ per unit of volume.
The respective volumes of the samples
to the distances of 3 kpc and 5 kpc used to derive
these functions were 
$1.22 \times 10^6$ pc$^3$ and $5.64 \times 10^6$ pc$^3$.
Errors of the PF's can be
obtained by scaling by the $1/\sqrt{N}$, where
$N$ are the numbers of systems in the respective period bins.
The period function derived from BW$_5$ can be used only above
$P \simeq 0.55$ day (or $\log P \simeq -0.25$) because distant, 
short-period, low-luminosity systems are 
eliminated from it by the magnitude 
limit of the OGLE survey at $I = 17.9$. The entries
of PF$_5$ which are affected by this selection effect
are taken in Table~\ref{tab4} in square brackets.
The period function are based on the apparent numbers of systems and
are not corrected for the systems missed because of the low orbital
inclination angles. 

\placefigure{fig2}

\placetable{tab4}   

\section{COLOR DISTRIBUTION}
\label{col}

The color distribution for the BW$_3$ sample in the $I$, $V-I$ system
was discussed in R97a. For comparison of the BW$_3$ and CL samples, 
the BW$_3$ data have been transformed here
to the $B-V$ color index. The BW$_5$ sample is not used in the
comparison because its color distribution is affected
by elimination of faint, red systems. Figure~\ref{fig3} shows 
a comparison of histograms representing the distributions
for BW$_3$ and CL. The agreement is not as close as in the case of
the period distributions. The two-distribution K--S test gives 
a 30 percent significance for rejection of the null 
hypothesis of identical distributions. 
The difference in the distributions
is caused mostly by the spike in the CL distribution 
at $(B-V)_0 = 0.75$.
However, the end points of the color distributions coincide 
well, with the distribution 
for BW$_3$ being a bit wider -- as expected --
because of the reddening correction uncertainties. 
In fact, four observational effects are expected to broaden 
the BW$_3$ color distribution; these are: 
(1)~uncertainties in the reddening in Baade's
Window, mostly from the spotty character of the reddening,
(2)~the crude model of the reddening adopted in
R97a, (3)~photometric blending of stars in the 
extremely dense BW field and (4)~the $(V-I)_0$ to $(B-V)_0$
color transformations.

\placefigure{fig3}

Because the BW$_3$ sample is expected to be complete and
statistically better defined than the CL sample, we will tolerate
the possibly larger color uncertainties for the BW$_3$ 
systems and use this sample in 
the next Section~\ref{cmd} as an external reference 
for comparison with individual clusters.
 The total range of colors observed for the BW$_3$ sample is 
$0.19 < (B-V)_0 < 1.54$. Because of the possibility
of errors for individual systems, we will also consider 
the 90 percent range, here defined as  the interval where 
90 from among 98 systems of the BW$_3$ sample 
are located. This range extends 
over $0.3 < (B-V)_0 < 1.2$ and almost perfectly coincides 
with the full range observed for all binaries of the CL sample,
which extends over $0.31 < (B-V)_0 < 1.21$. We note 
that while relatively red systems are seen in the BW$_3$ sample,
none of the stars in the CL sample is as red as CC~Com with
$(B-V)_0 = 1.24$ (Rucinski \markcite{ruc77} 1977); this
is partly expected as $M_V = 6.7$ of CC~Com is close to,
or perhaps even beyond the limiting levels of the 
cluster searches.

\section{COLOR -- MAGNITUDE DIAGRAMS}
\label{cmd}

The color -- magnitude diagram for the BW$_3$ sample is shown 
in Figure~\ref{fig4}. 
The thin lines in the figure give the observed 
isochrones for Praesepe and NGC~6791, which are used 
for reference. The former is a moderately old cluster 
with age about 0.9 Gyr while the latter is one of the oldest 
clusters known with age of about 6 -- 8 Gyr. Only one cluster 
in the CL sample is younger than Praesepe. It is Be~33, at 0.7 
Gyr. However, as was commented in Section~\ref{samples},
we are not sure if its only member, IK~Lyr, really belongs to it
so it has been decided to use Praesepe as a case of a ``young''
old open cluster with a contact system.

\placefigure{fig4}   

The band of the contact systems in Baade's Window in 
Figure~\ref{fig4} extends along the main sequence with a width of
about 1 magnitude, and shows a concentration of systems in the
region of the TOP of the oldest galactic disk population.
The width of the sequence may be due to 
observational errors and spots on the stars, but also to
a spread in the mass-ratios. The latter is entirely unaccounted 
in the absolute magnitude calibration, 
but its influence can be predicted 
by considering how total luminosity and total radiating 
area change with variation in the mass ratio. A small 
insert in the lower left 
corner illustrates how changes in the mass-ratio can modify 
the position of a contact system in the color--magnitude
diagram (for details, see CAL5). For identical stars 
($q=1$), the shift is upward by $-0.75$ mag, but 
for less massive secondary components 
($q \rightarrow 0$), the secondaries provide always relatively
more radiating area than luminosity, so that the color
becomes redder. The color shift is 
the largest for moderate mass-ratios around $q \simeq 0.5 - 0.6$.

\placefigure{fig5}  

\placefigure{fig6}

\placefigure{fig7}

Figures~\ref{fig5} -- \ref{fig7} show the positions of  
contact systems in the individual clusters. For each cluster, 
the approximate run of the respective observed isochrone is shown, 
together with positions of the isochrones for Praesepe and
NGC~6791. Close examination of the 
figures shows that the contact systems are {\it not\/} concentrated 
in the immediate vicinity of the respective TOP's. In fact, 
in those clusters where many systems were detected, such as Cr~261 or 
Be~39, the systems appear on both sides of each TOP, 
among the MS systems as well as among the Blue 
Stragglers. In all cases, except (marginally) for Praesepe 
and for Be~33 (where the association of  the only system 
to the cluster may be questioned), the locations of the 
TOP's themselves fall within the 90 percent range of the BW$_3$ 
sample, that is $0.3 < (B-V)_0 < 1.2$.
What we see is not that the positions of the systems are related
to the location of the TOP, but rather, that once the cluster TOP falls
into the above range, the systems can then appear anywhere within it.
Compare, for example, the diagrams for populous clusters such
as Cr~261, Be~39 or M~67. In the last case, only 3 systems are known,
but they span the whole width of the color range.

The last panel of Figure~\ref{fig7} contains the main sequence of
Praesepe with superimposed marks giving masses according to a 
$M_V$--mass calibration for disk stars by Kroupa et al.\ 
\markcite{kro93} (1993). As was discussed above, and as the
inserts in the figures illustrate, the unknown mass-ratios
can modify the luminosities and colors to some extent, but
we can expect that these would be the primary components 
which would define positions of the systems in the color--magnitude
diagrams. Thus, the last panel of Figure~\ref{fig7} gives a 
rough idea about the primary-component masses involved. They
are apparently concentrated in the range of about 0.65 -- 
1.6~M$_\odot$, with the maximum close to the 1~M$_\odot$. Thus, as has been 
known for some time, the contact binaries of the W~UMa-type are typically
composed of solar-type stars.

\section{PERIOD -- COLOR RELATION}
\label{PC}

The period--color (PC) relation is a useful tool 
for studies of contact binaries. Effectively, it is a relation 
similar to the color--magnitude diagram, but with one of the 
photometric parameters replaced by the orbital period, which 
is known with an accuracy several orders of magnitude higher 
than either brightness or color. The PC relation for 
the BW$_3$ sample (with $V-I$ as the base color) was presented 
in R97a, where the special significance of the short-period 
blue-envelope (SPBE) was also stressed. The concept
of the SPBE is similar to that of the Zero-Age main sequence, 
in the sense that a system can move only 
in certain directions away from the SPBE. Here, in the
period -- color plane, the directions are down and right 
(see Figure~\ref{fig8}). A system 
can be redder and larger (i.e.\ can have a longer 
orbital period) because of the evolutionary effects, 
while its color can be also redder because of the 
interstellar reddening. Location of the SPBE does depend 
on metallicity, and for low [Fe/H] it is shifted to bluer 
colors and shorter periods (CAL2).

\placefigure{fig8}

The PC relation for the BW$_3$ sample using the $(B-V)_0$ color 
is shown in Figure~\ref{fig8}. 
A new fitting formula for the SPBE which is over-plotted
in the figure was found by matching the previously used expression,
 $(V-I)_{SPBE} = 0.053 \times P^{-2.1}$, after its 
transformation into $B-V$ using the MS transformations
of Bessell \markcite{bes1} \markcite{bes2}
(1979, 1990). It is: $(B-V)_{SPBE} = 0.04 \times P^{-2.25}$. 
It must be stressed that the 
numerical values in both formulae have no physical significance. 
A few systems located slightly above the SPBE may 
be low-metallicity objects or cases of poor/blended photometry. 
What is unusual in Figure~\ref{fig8} is that 
we do not see a well-defined {\it period-color relation\/}, 
because the scatter is large, primarily due to the presence of some 
red, long-period systems filling the lower right of
the figure. While their locations are 
not a priori impossible, we do not see 
such systems in the sky-field sample (eg.\ the Hipparcos 
sample, CAL5), or in the CL sample
(see Figure~\ref{fig9}). We suspect that photometric blending 
of images leading to wrong colors or wrong
reddening corrections and/or period aliases may have
resulted in populating this part of the PC diagram. Inspection
of the OGLE light-curve data 
indicates that only two systems in this region
have well defined light curves, whereas most of the curves
show small amplitudes and a large photometric
scatter, which is possibly due
to the use of period aliases\footnote{The systems with
good light curves, but then possibly wrong colors, are 
BW3.053 and 7.147, while the low amplitude systems 
showing large light-curve scatter are:
BW3.022, 3.053, 5.075, 5.143, 5.157, 6.123, 7.112. Note that
3.053 and 7.112 are also among systems appearing
in the faint tail in the $M_V$ distribution in Figure~\ref{fig11}.}.

\placefigure{fig9}  

Because the SPBE is not expected to differ much within the range
of metallicities observed for the clusters of the CL sample,
all cluster systems are included in Figure~\ref{fig9}.
They have been 
divided into two groups to avoid congestion of the symbols. 
The four oldest, most populous clusters are shown in the 
upper panel of the figure, while the younger ones are 
collected in the lower panel. We see signatures of low 
metallicity for the systems in Tom~2 and Be~33, but the 
rest conform to the expected tendency of confinement 
below the SPBE for normal-metallicity contact systems.
Open symbols signify Poor Thermal Contact systems. Two among
them, one in Cr~261 and one in NGC~188, are clearly more evolved,
showing longer orbital periods than systems of similar colors.

\section{AMPLITUDES OF LIGHT VARIATIONS}
\label{ampl}

As was discussed in Section~5 of R97b, distributions of the light-curve 
amplitudes contain information about the 
mass-ratio distribution. Because the light curves are 
dominated by geometrical effects of the strong distortion 
of the components, rather than by properties of 
stellar atmospheres (such as limb and gravity darkening laws), 
it is relatively easy to predict distributions of the 
amplitudes of light variations assuming random orbital inclinations 
and some plausible mass-ratio distributions. 
When mass-ratios are large ($q \rightarrow 1$), large 
and small amplitudes can be observed, depending on the orbital
inclination, but when the mass-ratios are small, only small 
amplitudes are possible, irrespectively of the inclination. 
However, the inverse problem of {\it determination\/} of the 
mass-ratio distribution, $Q(q)$, from the amplitude 
distribution, $A(a)$, is not an easy one as it would
involve a solution of 
an integral equation representing a convolution of distributions
(R97b). Such a determination could be contemplated for a sample of the 
order of one thousand objects or more.

\placefigure{fig10} 

Here, we limit ourselves to a 
comparison of the amplitude distributions for the BW$_3$ and CL 
samples which is shown in Figure~\ref{fig10}. The
distributions are apparently slightly 
different in that the CL sample appears to contain more 
large-amplitude systems than the BW$_3$ sample. 
The numbers per bin are small so the differences
are not really significant. Besides, the difference can be explained
by the use of the $I$-band amplitudes for the BW$_3$ sample
which are expected to be systematically slightly smaller 
that the amplitudes in the $V$ band. A simple
scaling would not be prudent as conversions depend on 
combinations of geometrical parameters, but the effect is
not expected to be larger than 3 to 8 percent. Thus, taking
into account the per-bin uncertainties, we conclude that the
amplitude distributions for the BW$_3$ and CL samples have 
basically identical shapes. However, as can be seen in Figure~\ref{fig10},
both distributions are very different from
that for the bright systems of the sky field (R97b), the latter
being heavily biased by the large-amplitude systems which tended
to be preferentially detected in non-systematic searches of the sky.

In the paper on Cr 261, Mazur et al. \markcite{maz95}
(1995) pointed out an interesting property of those
contact systems which occur among 
the blue stragglers of the cluster: All of them were found to have 
small amplitudes, which raises a possibility that all of these 
systems have small mass-ratios. A meaningful analysis 
of the above effect could be done only for those clusters which have 
contact systems on both sides of the TOP. 
The amplitudes have been shown schematically
for the clusters of the CL sample in 
Figures~\ref{fig5} -- \ref{fig7}. There exist some differences
between individual clusters, eg.\ in NGC~6791, all systems have
small amplitudes, while all in NGC~188 have large amplitudes,
but these may be due to small number statistics. Generally, we do not
see any clear tendency for small amplitudes above the 
respective TOP's and Cr~261 remains the only cluster
where the effect is rather clearly visible (the two PTC systems
excepted). The tendency is not so obvious in Be~39 which is the 
next cluster in terms of the number of contact systems. As expected,
the BW$_3$ sample does not show any segregation in the 
amplitudes along the main sequence, but this can be 
explained by a mixture of ages in  the BW$_3$ sample. 
The amplitudes for that 
sample have been shown symbolically in Figure~\ref{fig4}.

\section{LUMINOSITY FUNCTION}
\label{LF}

The absolute magnitudes $M_V$ for the BW and CL samples come
from two entirely different determinations. The 
ones for the BW sample have been estimated via the 
$M_I = M_I (\log P, (V-I)_0)$ calibration in R97a (this 
involved an iteration in $E_{V-I}$), and then adjusted via
$M_V = M_I + (V-I)_0$; the ones for the CL sample result 
from the observed magnitudes $V$ and assumed 
distance moduli of the clusters. In spite of 
coming from very different sources, the $M_V$ absolute magnitude
distributions turn out to be again similar, as can be seen 
in Figure~\ref{fig11}, in that both show
maxima in the interval $3 < M_V < 6$. The CL sample
shows some deficiency at the faint end relative to BW$_3$, but
is remarkably similar to BW$_5$ which we know to be more affected
by the magnitude limit of the OGLE survey than BW$_3$. For the
limiting magnitude of the OGLE search of $I_{lim} = 17.9$,
taking into account the interstellar extinction and the typical
colors, as they vary along the absolute-magnitude sequence,
the expected limits for BW$_3$ and BW$_5$ are $M_V \simeq 5.5$
and $M_V \simeq 4.2$, respectively.
Nominally, the CL sample should for some of the clusters 
reach depths of $M_V 
\simeq 6-7$, that is even deeper than the BW$_3$ sample; however,
its low luminosity limit is, by necessity, 
a rather fuzzy one, being defined by 
the increases in the photometric errors for the contributing clusters rather 
than by a fixed distance, as in the 
case of the BW samples. Because of this deficiency, the CL sample is not 
considered in the discussion of the luminosity function. 
We have a good reasons to think that the BW$_3$ sample 
is fully complete to $M_V \simeq 5.5$, as the star number densities 
estimated to 3 kpc were found in R97a 
to be identical to those for 2 kpc. 
A few faint systems that populate the very 
tail of the BW distribution in Figure~\ref{fig11}
to $M_V \simeq 9$ must be nearby objects.
They have been checked in the OGLE data 
for anomalies, but appear to have well 
defined light curves (but errors in colors 
are obviously possible); these are 
the variables BW3.053, 4.040 5.114, 7.112, 8.072. 
All of them, except BW7.112, have well defined light curve with
large amplitudes.

\placefigure{fig11}  

The absolute magnitude distributions shown in Figure~\ref{fig11} have been 
converted into the luminosity functions (LF's) by simply dividing 
the system numbers by the total volumes of the BW$_3$ 
and BW$_5$ samples, $1.22 \times 10^6$ pc$^3$ and $5.64 \times 10^6$ 
pc$^3$, for the depths of 3 kpc and 5 kpc, and for the $40' \times 40'$
field of view. The limiting absolute magnitudes for the BW$_3$ and BW$_5$ 
samples (assuming constant interstellar absorption beyond 2 kpc,
see R97a) are $M_V = 5.5$ and 4.2, 
respectively. Beyond these completeness limits, 
the numbers of stars are expected to 
decrease because of the shrinking search 
volumes. These decreases should follow the standard 4-times per magnitude 
volume-size relation and thus can be 
accounted for by the volume corrections. 
Obviously, an application of such corrections magnifies the increasing 
Poisson errors so that an extension to fainter magnitudes can be done only 
slightly beyond the completeness limits of the survey. In our case, this 
extension was made into only two or three bins 
beyond the respective completeness limits of the BW samples.

\placetable{tab5}

The luminosity functions derived for the total volumes of the
BW$_3$ and BW$_5$ samples are listed in Table~\ref{tab5}.
In this table, $N_3$ and $N_5$ are the numbers of contact systems
in the BW$_3$ and BW$_5$ samples in one magnitude wide bins,
centered on $M_V$. LF$_3^{obs}$ and 
LF$_5^{obs}$ are the corresponding 
observed luminosity functions, in units
of $10^{-5}$ pc$^{-3}$, with entries which have been corrected
by the volume correction of 3.981 times per one magnitude increment
taken in square brackets. 

\placefigure{fig12}

In addition to the observed luminosity functions
for the BW$_3$ and BW$_5$ samples, Figure~\ref{fig12} shows
the luminosity function for the solar neighborhood MS stars LF$_{MS}$
(Wielen et al.\ \markcite{wie83} 1983), which has been
arbitrarily scaled down by a factor of 130. This factor
was selected to approximately match both LF$_{BW}^{obs}$ in the interval
$3 < M_V < 5$ where the statistics should be the most
reliable, i.e.\ it should not be affected by small-number
fluctuations at the bright side and discovery-selection effects
at the faint side. Obviously, the same factor of 130 gives the inverse
apparent frequency for the W~UMa-type systems and directly
shows that these binaries are indeed very common. We discuss the
frequency of occurrence more fully in the next section.

A comparison of the luminosity functions in Figure~\ref{fig12} reveals 
surprisingly close similarities between the contact binary and MS
functions (note in particular the dip at 
$M_V = 7$ for BW$_3$). However, there 
exist also some obvious differences between the shapes of
LF$_{BW}$ and LF$_{MS}$; in particular, we
see relatively fewer high-luminosity systems than low-luminosity ones,
an effect which is stronger for the BW$_5$ sample.
These differences can be ascribed to the fact 
that LF$_{MS}$ is based on the local volume defined by the distance of 
less than 20 pc from the Sun, while the 
BW functions were obtained from a pencil-beam search reaching deep into the 
galactic space. If the contact binaries follow the distribution of
disk stars, then we can expect changes in their numbers due to the
structure of the galactic disk. This links the luminosity function
and frequency of occurrence of contact binaries with
the description of the galactic disk structure.

\section{FREQUENCY OF OCCURRENCE OF CONTACT SYSTEMS}
\label{freq}

\subsection{Influence of the galactic disk structure}

The models of Bahcall \& Soneira \markcite{bah81} (1981) were used by 
Paczynski et al.\ \markcite{pac94} (1994) to find out the numbers 
of disk stars along the OGLE line of sight. 
The same approach has been followed here with 
some minor modifications. The ratio of the star number density 
$n(d)$ at a distance $d$ 
to the local density $n_0$ can be expressed as a product of two 
exponentials: 
$c(d) = n(d)/n_0 = \exp(d/h_R)\, \exp(-|z|/h_z)$, 
where $h_R$ and $h_z$ are the 
galactic disk length and height scales, respectively. $h_z$ depends on the 
absolute magnitude of the MS stars as the galactic-plane concentration
is spectral-type dependent.
Bahcall \markcite{bah86} (1986) suggested the following relations: $h_z = 
90$ pc for $M_V < 2.3$ and $h_z = 325$ pc for $M_V > 5.1$, with a linear 
interpolation between these values: 
$h_z = 90 + 83.9 \, (M_V - 2.3)$ pc. Paczynski 
et al.\ \markcite{pac94} (1994) noted 
that for the galactic coordinates of the OGLE 
search ($b \simeq -4^\circ$, 
implying $z \simeq 0.068\,d$) and for $h_R = 3.5$ 
kpc, the two exponential terms practically 
cancel out and the density stays approximately constant. 
However, the newest discussion of the galactic disk by 
Sackett \markcite{sac97} 
(1997) suggests a shorter disk length scale, 
$2.5 < h_R < 3.0$ kpc, so that the 
planar term may win leading to an increase in the numbers of stars at large 
distances. Since, as we argued in R97a, the contact binaries are apparently 
genuine members of the old disk population, 
the increase in star numbers along 
the OGLE line of sight for the shorter $h_R$
could possibly explain the high numbers of contact 
binaries in the BW sample. The density change factors $c(d) = 
n(d)/n_0$ are shown in 
Figure~\ref{fig13} for two values of $h_R$, 2.5 and 3.5 kpc. 
The shorter value of $h_R$ results in a larger differentiation between
the star number densities for various values of $M_V$. The
decrease in numbers of early-type systems 
with distance is clearly visible for both values of $h_R$.

\placefigure{fig13} 

Comparison of the luminosity function for contact binaries 
with the local MS function requires knowledge of the mean weighted values 
of $c(d)$ to the limits of 3 and 5 kpc, obtained by taking into account the 
increasing volume with the distance. Because of this weighting, 
the systems at large 
distances contribute more to the mean densities obtained from the BW samples 
than the local systems. 
The weighted values of the factors $c(d)$ for each 
bin of $M_V$ can be calculated from: $\bar{c}(d_l, h_R, h_z(M_V)) = 
\int_0^{d_l} c(\rho)\,\rho^2\,d\rho / \int_0^{d_l} \rho^2\,d\rho $.
The luminosity functions can now be
related through LF$_{BW}^{corr}$ = LF$_{BW}^{obs} / \bar{c}(d_l, h_R, h_z)$,
and LF$_{BW}^{corr}$ = LF$_{MS} \times f$,
where $f$ is the frequency of occurrence of contact binaries.
$d_l$ is the depth of the sample equal to respectively 3 or 5 kpc, while
$h_R$ have been assumed to be equal to 2.5 and 3.5 kpc;
$h_z(M_V)$ is given by the interpolation formula of
Bahcall \markcite{bah86} (1986) cited above.
The corrected luminosity functions are given in Table~\ref{tab6}.
 
\placetable{tab6}

The above formulation permits a comparison of the LF's for the contact 
binaries and the MS stars on the per $M_V$-bin basis.
The frequencies $f$ derived in such a way are shown 
in Figure~\ref{fig14} and are also listed in Table~\ref{tab7}.
The table gives the inverse apparent frequency of occurrence of contact
binaries, $1/f$, expressed as the number of MS
stars per one contact binary. The line
WM gives the weighted mean values of the inverse frequencies
over the available $M_V$. Because of the large errors in the
first bin at $M_V = 1$, there are no
changes in these frequencies if this bin
is excluded from averaging. This bin is in fact affected
by the bright limit of the OGLE survey. The 
bright limit of the sample at $I \simeq 14.1$ (R97a) translates,
for the average interstellar absorption in this direction, 
into the absolute magnitude limits for BW$_3$ and BW$_5$ of $M_I \simeq
0.8$ and $-0.3$, respectively. Taking into account
typical colors along the contact binary sequence, these
upper limits correspond to $M_V \simeq +1.0$ and $-0.1$. Thus, almost
nothing can be said about the frequency of occurrence
for $M_V < 1.5$. As we will see below in Section~\ref{bright},
the sky sample of bright systems gives us information that
the frequency of contact binaries must fall down for
such systems.

\placefigure{fig14}

\placetable{tab7}

Except for the uncertainty at the 
bright end, Figure~\ref{fig14} shows that the corrections
for the galactic structure produce the frequency distributions which are 
remarkably flat. The systematic differences in the 
luminosity function in 
Figure~\ref{fig12}, which are best visible for intrinsically 
bright systems, are taken into account by the $M_V$-dependence of the 
disk height scale. It is possible that the
corrections are too large for the bin at $M_V =2$, but the
data are consistent with the flat frequency distribution even for this
bin. The resulting apparent frequency of the contact binaries 
in the BW direction is one system 
per about 130 MS stars for $h_R = 2.5$ kpc 
and one system per about 100 MS stars for $h_R = 3.5$ kpc. Unfortunately,
at this stage, we cannot decide which number is the correct one. On one
hand, uncertainties in the galactic disk structure are too large to make
a preference with respect to $h_R$. On the other hand, we cannot use
an argument that the contact binary frequency should have a flat
distribution in $M_V$ to determine $h_R$. Although our BW samples
are perhaps among the first volume-limited ones in this particular
galactic direction, we would not like to over-interpret the results
as we feel that our crude interstellar-absorption
model (R97a) may couple with the inferred spatial distribution
of contact systems along the OGLE line of sight.

\subsection{Frequency 1/130 or 1/100; why so high?}

The frequency of occurrence of contact binaries in the BW samples
of 1/130 or 1/100 that we have estimated above is some two times
higher than the previous estimate of 1/250 -- 1/300 that we
obtained from the same BW material in R97a and for the old open
clusters in CAL1. We will try to 
find explanations for these discrepancies in turn.

First of all, we should stress that the space density of contact 
binaries derived in R97a for BW$_3$, to $M_I = 4.5$ 
(or equivalently to $M_V 
\simeq 5.5$) of $7.6 \times 10^{-5}$ systems per pc$^3$ is a correct one. 
Paradoxically, the problem is with relating this number to the number of MS 
stars in the same volume. Since the numbers of stars that had been analyzed 
for variability by the OGLE project in successive apparent magnitude 
bins were not available, the numbers of stars with good photometry 
were used in R97a instead. For fainter magnitudes, the 
quality of photometry drops and the blending becomes more severe.
We made therefore an assumption that the OGLE sample of stars with good 
photometry and the sample analyzed for variability had similar 
biases. The counting corrections for the good-photometry 
sample were quite large for fainter magnitudes, of the order 
of a factor of 2 or more, and this could be a source of a potentially
large error. It is quite possible that 
the assumption of the identical character of biases in both samples was 
incorrect. The approach presented here is simpler: We find 
the BW luminosity function by simply
counting the numbers of the contact systems, 
then correct it for the galactic disk 
structure and compare it with that for the MS stars from 
Wielen et al.\ \markcite{wie83} (1983) by taking the ratio
of the functions. If, for some
reason, some contact systems are missed, we 
can only under-estimate their frequency of occurrence. As we
discussed above, the main difficulty here is our insufficient 
knowledge of the disk length scale $h_R$, but the frequency comes
out large for any of the two possible choices. Thus, we feel that
the frequency for the BW sample is indeed high, about two times 
higher than estimated in R97a.

Concerning the frequencies observed in old open clusters:
An earlier preliminary estimate of the apparent frequency 
in the clusters (CAL1) gave one contact system per $275 \pm 75$
MS stars. This estimate was based on seven clusters of
considerable spread in age from among the eleven that contribute
to the present CL sample. As we know, we have good reasons
to suspect that numbers of contact binaries increase with time.
Therefore, the difference between the above estimate 
and the new determination for the BW sample
may indicate an older -- on the average -- age of the latter 
sample. Estimates of the apparent frequency were published for
for two of the four clusters studied subsequently to CAL1, Cr~261
(Mazur et al.\ \markcite{maz95} 1995) and NGC~7789 
(Jahn et al.\ \markcite{jah95} 1995). Depending how the
cluster membership of the systems is established, the frequencies 
were found to be 
1/140 -- 1/88 for Cr~261 and 1/178 -- 1/150 for NGC~7789.
These determinations are in full
agreement with our new value for the BW sample and with the 
frequency showing an increase with age because
Cr~261 is an older cluster than NGC~7789.

\subsection{Is there a low-luminosity end of the contact binary sequence?}

The result on the high apparent frequency of incidence of
contact systems in the BW sample is based primarily on 
the moderately bright among them, mostly within  $2.5 < M_V < 5.5$. 
The volume corrections are large for systems fainter than $M_V = 5.5$, 
and there are no data for $M_V > 7.5$. However, we have no basis 
to assume that 
the contact binary sequence stops at $M_V \simeq 7.5$ because
we appear to see a few even fainter systems in the BW$_3$ sample
(provided these are not artifacts of too red colors).
Probably the only good argument for the existence of the sudden drop
in the sequence is the constancy of the star number density when
the limiting depth of the BW sample 
is evaluated for the depths of 3 and 2 kpc 
(R97a); apparently, no intrinsically faint systems (whose
existence would produce a density increase for the smaller volume)
have been detected in the solar neighborhood. 
However, the 2 kpc sample consist of only 
27 objects, so that this argument is 
not very strong. Otherwise, the fact that 
we do not see faint contact systems 
may fall into the class of the {\it absence
of evidence\/} versus {\it evidence of absence\/} reasoning.
We do see a sharp end of the period distribution for both, 
BW$_3$ and CL, samples at about 0.225 day, but this is not 
an argument for a sharp cutoff in the absolute magnitude 
sequence as $M_V$ changes very rapidly for short 
periods and red colors, where 
the period-color relation becomes almost 
vertical (see Figures~\ref{fig8} and 
\ref{fig9}). The period distribution is stretched for short periods 
when logarithmic units are used (see Figure~\ref{fig1}) and 
there is more room for short-period systems. Arguably, the logarithmic
units are the proper ones in view of the
power-law dependencies governing the angular momentum loss.

The predictions based on the full convection limit 
(Rucinski \markcite{ruc92} 1992) place the
expected low luminosity limit of the contact-binary
sequence at $B-V \simeq 1.5 - 1.6$, 
that is at spectral types M2 -- M4, leaving a large gap in the 
parameter space between the location of current ``record holder'', 
CC~Com\footnote{CC~Coma, with its $M_V = 6.7$ determined
from the combined photometric and spectroscopic study
of Rucinski et al.\ \markcite{rww77} (1977) follows perfectly,
to within 0.1 mag., the Hipparcos calibration CAL5. This is
an argument that the
calibration can be used for intrinsically faint systems.} 
at $(B-V)_0 = 1.24$ and $P = 0.221$ day 
and the expected full-convection limit. 
We note that the close pair of M-type dwarfs, BW3.038
(Maceroni \& Rucinski \markcite{mac97} 1997) with the
period of 0.1984 day, is on its way to becoming a contact
system. Perhaps corresponding contact systems already exist
and we simply have not found them? The fact that such faint systems
have not been detected in the sky field is not an argument
as the sky has been searched very poorly and unsystematically.

\subsection{Comparison with the sky-field and cluster data}
\label{bright}

The variability amplitude distribution (Section~\ref{ampl}),
which is apparently biased to large values of the amplitudes, gives 
us a strong indication that many low inclination systems remain to 
be discovered in the solar neighborhood. The pioneering
study of Duerbeck 
\markcite{due84} (1984) attempted to correct for the 
orbital inclination discovery biases in the sky sample,
leading to an estimate of the apparent frequency of 
occurrence of contact binaries of one system per about 
one thousand MS stars. In view of the subsequent work (CAL1, R97a),
this estimate seemed too low by a factor of 3 -- 4 times. Now, a new 
increase in the apparent frequency is postulated to the level of
one contact system per about one hundred MS stars, a change
which may be considered as quite drastic. 
Therefore, we must inquire whether 
the current results based on the BW sample 
are {\it consistent\/} with the sky field statistics. 

With the luminosity functions in Table~\ref{tab6} 
or in Figure~\ref{fig12}, 
one can easily calculate the number 
of stars in the ``contact binary sky'' by 
considering the space volume accessible for discoveries
for a given limiting magnitude $V_{lim}$. We can use either the
luminosity functions LF$_3^{corr}$ or LF$_5^{corr}$ 
or the scaled main-sequence function,
recognizing that the latter has a smaller statistical uncertainties.
Given a LF~($M_V$), as in Table~\ref{tab5}, one can calculate 
for each bin of 
$M_V$ the total number of stars observable to a given apparent limiting 
magnitude $V_{lim}$ from: 
$n (M_V, V_{lim}) = {\rm LF}(M_V) \times 4/3 \pi d^3(M_V, V_{lim})$,
where $d(M_V, V_{lim}) = 10^{1+0.2(V_{lim}-M_V)}$. The examples 
for $V_{lim} = 7.5$ (left vertical axis)
and $V_{lim} = 12.5$ (right vertical axis)
are shown in Figure~\ref{fig15} for the MS luminosity function
scaled by the factor of 130; for other
$V_{lim}$ the numbers can be obtained by the usual uniform spatial
density scaling ($10^3$ times per a five magnitude
difference or 3.981 times per one magnitude). The last column of
Table~\ref{tab5} gives $n (M_V)$ for $V_{lim} = 7.5$.

\placefigure{fig15}

How do the results compare with the data for 
the sky field? The predicted numbers of the faint end
of contact systems are low, but -- still -- 
of the order of one hundred faint systems similar to CC~Com 
are expected over the whole sky to
$V_{lim} = 12.5$, in contrast to a dozen or so currently known.
While there is no question that substantial contributions for 
a resolution of this discrepancy should 
come from large scale, yet simple surveys of the sky, 
similar to that currently conducted by Pojmanski \markcite{poj97} 
\markcite{poj98} (1997, 1998), a survey similar to OGLE, but deeper
would probably offer a more efficient approach to learn about
the faint end of the sequence. If the limit of OGLE
were not $I_{lim} = 17.9$, but 19.9, we would already know the
luminosity function beyond the position of CC~Com.

The situation is very different at the bright end. 
The BW samples give us practically no information
as only one, the same, contact system appears in both BW samples 
in the $0.5 < M_V < 1.5$ bin, so that the sequence really starts 
with the bin $1.5 < M_V < 2.5$. This dearth of the systems
is due to the bright limit of the OGLE survey at 
$M_I \simeq 0.8$ and $-0.3$, for the BW$_3$ and BW$_5$ samples,
respectively. Thus, we are forced to use the scaled MS data at the bright
end and then check if the frequency scaling does apply here.
We should note at this point that the CL sample has
a bright end which is
defined by the two brightest systems, HQ~Mus and V732~Cas, at $M_V = 2.0$. 

Figure~\ref{fig15}
shows the well known fact that the visibility of stars in the
sky is heavily biased toward intrinsically bright objects.
The predictions based on the scaled MS data 
give $40 \pm 7$ contact systems in the whole sky to $V_{lim} = 7.5$, 
but in that number as many as
$23 \pm 5$ would be contributed by the bin
$1.5 < M_V < 2.5$. The additional $10 \pm 2$ systems would come from
the next bin $2.5 < M_V < 3.5$ (see the last column
of Table~\ref{tab5}). If we eliminate the first bin
at $M_V =2$, the total number of systems
with $M_V > 2.5$ should be $17 \pm 2$. These predictions, when
confronted with the observed numbers of contact binaries in the
sky directly tell us that their frequency must decrease at high
luminosities as we simply do not see that many bright contact
binaries. At present, we know of one system 
at $V = 4.7$ ($\epsilon$ CrA), one system at $V = 5.9$ ($44i$ Boo~B) 
and six further systems (S~Ant, V535~Ara, RR~Cen, VW~Cep, AW~UMa and
HT~Vir) are brighter than $V_{lim} = 7.5$. We know from the
Hipparcos survey (CAL5, Duerbeck \markcite{due97} 1997)
that most of them are indeed
intrinsically luminous: Among those 8 systems, 6 fall in the interval
$1.5 < M_V < 3$, while two ($44i$ Boo~B and VW~Cep) have $M_V > 5$. 
Thus, we see about one half of the number of
systems predicted by the high
frequency of occurrence of 1/130, but most of this discrepancy comes
from the high luminosity end where the frequency must be definitely
lower.

Let us assume that the number of contact binaries to 
$V_{lim} = 7.5$ is indeed 8. We can learn about the discovery selection
effects at fainter magnitudes by considering the numbers of systems
predicted to various $V_{lim}$. For each magnitude 
increase, we expect an increase in the numbers of contact binaries 
by 3.981. Then, the sequence for the progression in
the magnitude limits, $V_{lim} = 
7.5, 8.5, 9.5, 10.5, 11.5, 12.5$, should lead
to the predicted numbers of the systems to be 8, 32, 127, 505,
2009, 8000. Since we know some 600 
contact binaries in the sky (some fraction of that in localized 
deep-search areas), we have a direct indication of 
discovery selection effects appearing at 
the level of about $V_{lim} \simeq 10 - 11$.  
Only wide field surveys can confirm or disprove this conjecture.

\subsection{Comparison of the frequency of 
contact binaries to that of other MS binaries}

When comparing the contact systems with 
other binaries we must remember that the former are located at the 
very end of the angular momentum and period sequences 
and we do not necessarily 
expect a perfect continuity over the whole range of orbital periods 
spanning several orders of magnitude. The currently best 
data on the period 
distribution for MS binaries are those by 
Duquennoy \& Mayor \markcite{duq91} 
(1991) who found that the distribution can be approximated, 
in the logarithm of the period, by a wide Gaussian with a 
maximum at $\log P = 4.8$ and $\sigma \log P = 2.3$, 
with the period $P$ expressed in days. Various techniques
contributed to this result and the normalization of the
distribution is somewhat uncertain. 
The recent results on the binarity of solar-type 
stars in the range where this distribution has a maximum 
(5 -- 50 AU) offer a way of relatively reliable normalization 
of the distribution in the sense of spatial frequency 
(i.e.\ the frequency free of geometrical effects of 
unknown inclination). Patience et al.\ \markcite{pat98} (1998) 
found that in the range of orbital periods, 
$3.7 < \log P < 5.2$, the frequency of incidence is 
$0.14 \pm 0.03$ (this means that one among about 7 solar-type stars 
is a binary with a period in the range
14 -- 430 years). This normalization 
has been used in Figure~\ref{fig16}. In plotting the contact 
systems, it has been assumed that their total spatial frequency 
of occurrence is 1/80 which was (somewhat conservatively)
estimated to correspond to the apparent frequency of 1/130.

\placefigure{fig16}

We clearly see in Figure~\ref{fig16} that the contact binaries
with periods {\it shorter\/} than 0.6 -- 0.7 day are very common forming
a sharp peak extending well above the main-sequence relation.
However, we should note the significant under-representation of contact
systems with periods {\it longer\/} than 0.6 -- 0.7 day. The latter
can be detected to very large distances and are present in
the OGLE sample, but they are very rare in terms of
the spatial density. One would expect that, in
a diagram like Figure~\ref{fig16}, their place is occupied
by close, short-period, detached binaries
which have not yet lost enough angular momentum to enter into
direct contact; however, no statistics similar to that
available for contact binaries in Baade's Window exists for 
such systems. Any attempts to find a trough in the period distribution
on the long-period side of the contact-binary peak will obviously
confront very different discovery selection effects for detached
and contact eclipsing systems.

\section{CONCLUSIONS}
\label{concl}

The main conclusion of this paper is that the
two samples of disk population contact binaries
give very similar results in almost every respect, in spite
of very different origins of the samples. The
similarities are observed in practically
all distributions: those of the orbital periods and colors, 
of the luminosity functions and of the variability
amplitudes. This is surprising and
unexpected, as many observational effects would tend to 
make the distributions different. The CL sample 
would be expected to be particularly inhomogeneous
as it was obtained by combining data for 11 
different clusters ranging in age roughly by an order of 
magnitude, within 0.7 to 7 Gyr. Selection of these clusters 
was not systematic. In addition, they were observed to different 
limiting magnitudes, with various equipment and differing 
search areas. If any mass segregation 
would take place in a cluster, the CL sample should 
contain preferentially more massive systems, as typically
only central parts of the clusters are only observed.
Thus, we conclude that in spite of the 
small statistics -- and possibly partly by coincidence -- 
the available mixture of 11 clusters in the CL sample has been 
representative in the sense that it 
has not introduced its own observational biases. 
Thus, it would be hard to avoid a conclusion that it is 
the formation process of the contact systems 
which creates those same distributions irrespectively of 
the age of the population. 

It has been found that contact systems typically appear in the 
period interval $0.23 < P < 0.7$ day and the color
interval $0.3 < (B-V)_0 < 1.2$. The turn-off points (TOP) 
of the clusters forming the CL sample all fall into the
same color interval. However,
the systems do not appear close to the respective TOP's, but
can appear anywhere in the above color range. 

By comparing the galactic-disk corrected
luminosity function derived from the BW sample with that
for the MS stars in the solar neighborhood by Wielen et
al. \markcite{wie83} (1983), the apparent frequency of
occurrence of contact systems in the interval $2.5 < M_V < 7.5$ was
found to be surprisingly high, at one contact system per
about 130 main sequence stars of a given absolute magnitude
for the galactic disk exponential length-scale $h_R = 2.5$ kpc; the
contact binaries would be even more common, with the apparent
frequency of one per about 100 MS stars for a longer scale of
$h_R = 3.5$ kpc. This high frequency is observed in the oldest
among the open clusters such as Cr~261 or NGC~188, 
which suggests an advanced age of the BW sample systems. 
Total absence of contact systems in clusters younger
than 0.7 Gyr and their low numbers in clusters younger than
about 2 Gyr suggest a frequency of occurrence strongly dependent
on the age of the stellar system. The reduction 
in the frequency for younger ages
is difficult to quantify due to
the low numbers of systems involved. The frequency of occurrence is
at present the only property which appears to be different for the
contact systems in old open clusters and in Baade's Window.

The BW luminosity function determinations
suffer from large search-volume corrections for
$M_V > 5.5$, but they do cast doubt on the location of the faint end
to the contact binary sequence, currently defined by
the K5-type system CC~Com at $M_V = 6.7$. The BW data
do not extend above $M_V \simeq +1.5$ so that the luminosity
functions and frequencies of occurrence cannot be determined
for intrinsically bright systems. However, the sky-field 
sample of bright stars to $V_{lim} = 7.5$, which 
presumably has been fully screened for the presence of
contact systems, indicates a clear
decrease in the apparent frequency for the high luminosity
systems. This decrease cannot be quantified due to the small
number statistics for the bright systems in the sky-field sample; 
a factor of two or three drop at $M_V = 2$ is quite likely with
much larger reductions for still brighter systems. 

The limitations of this work are twofold:
Because we are uncertain about the completeness for faint
systems, we have not directly addressed 
the matter of the mass distribution of the contact systems,
but this can be roughly estimated from the available $M_V$
distributions and the diagram in the last panel in 
Figure~\ref{fig7}. Also, since we have no idea about the
mass-ratios of individual systems, we cannot say anything
about the orbital angular momenta, which are dominated
by the mass-ratio dependent term in: $H \propto M^{5/3} P^{1/3}
\frac{q}{(1+q)^2}$.

This work does not include 
Population~II contact systems of the type recently
found in large numbers among blue stragglers of globular
clusters (for most recent references, see Mateo \markcite{mat96}
(1996) and the new discoveries in $\omega$~Cen and M4
by Ka\l u\.zny et al.\  (1997a, 1997b, 1997c)). However,
halo-population stars are exceedingly rare in the solar
vicinity, at the level of 0.125 - 0.15 percent of
all stars (Bahcall \markcite{bah86}
1986, Reid \& Majewski \markcite{rei93} 1993), so that
no contact systems, even at high frequency of occurrence, would be
expected among 98 members of our basic sample BW$_3$.
Thus, the results presented here are relevant solely to the
most common contact binaries of the galactic disk
field and of the old open clusters.

\acknowledgements
This work is dedicated to Janusz Ka\l u\.zny, 
my friend and colleague for over 20 years. Without his hard work, this
study would have been entirely impossible.

\bigskip
Special thanks are due to Carla Maceroni and Hilmar Duerbeck for extensive
and useful suggestions and comments on the first version of the paper.

\newpage
 
\centerline{Figure captions:}

\medskip
 
\figcaption[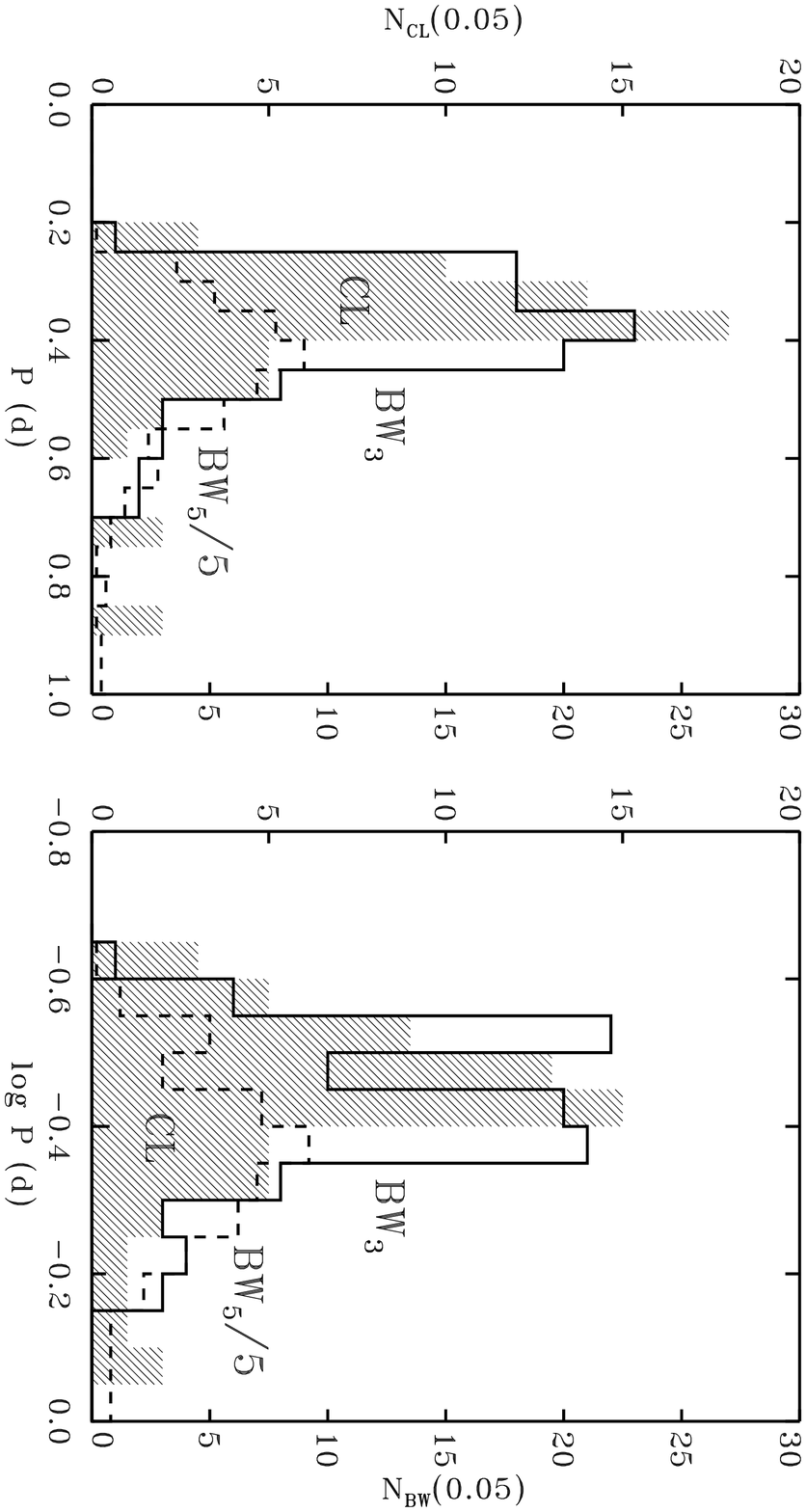] {\label{fig1} 
The period distributions for the Baade's Window 3 kpc
(BW$_3$, continuous line), 5 kpc (BW$_5$, broken line), 
and open cluster (CL, hatched) 
samples, binned in linear (left panel) and logarithmic
(right panel) units of the orbital period expressed
in days. The histograms in this and the following
figures are not normalized in order to visualize directly
the numbers of systems in each bin and thus permit to
judge the Poissonian errors in the distributions.
Note that the left (CL) and right (BW) side
vertical axes in this and the following figures are in
proportion 3:2, to account approximately for the different
sizes of the BW$_3$ and CL samples of 98 and 63 systems.
The data for 238 members of BW$_5$ are shown scaled by 5 times, but 
the volume of BW$_5$ is actually 4.6 times larger than that
of BW$_3$.
}

\figcaption[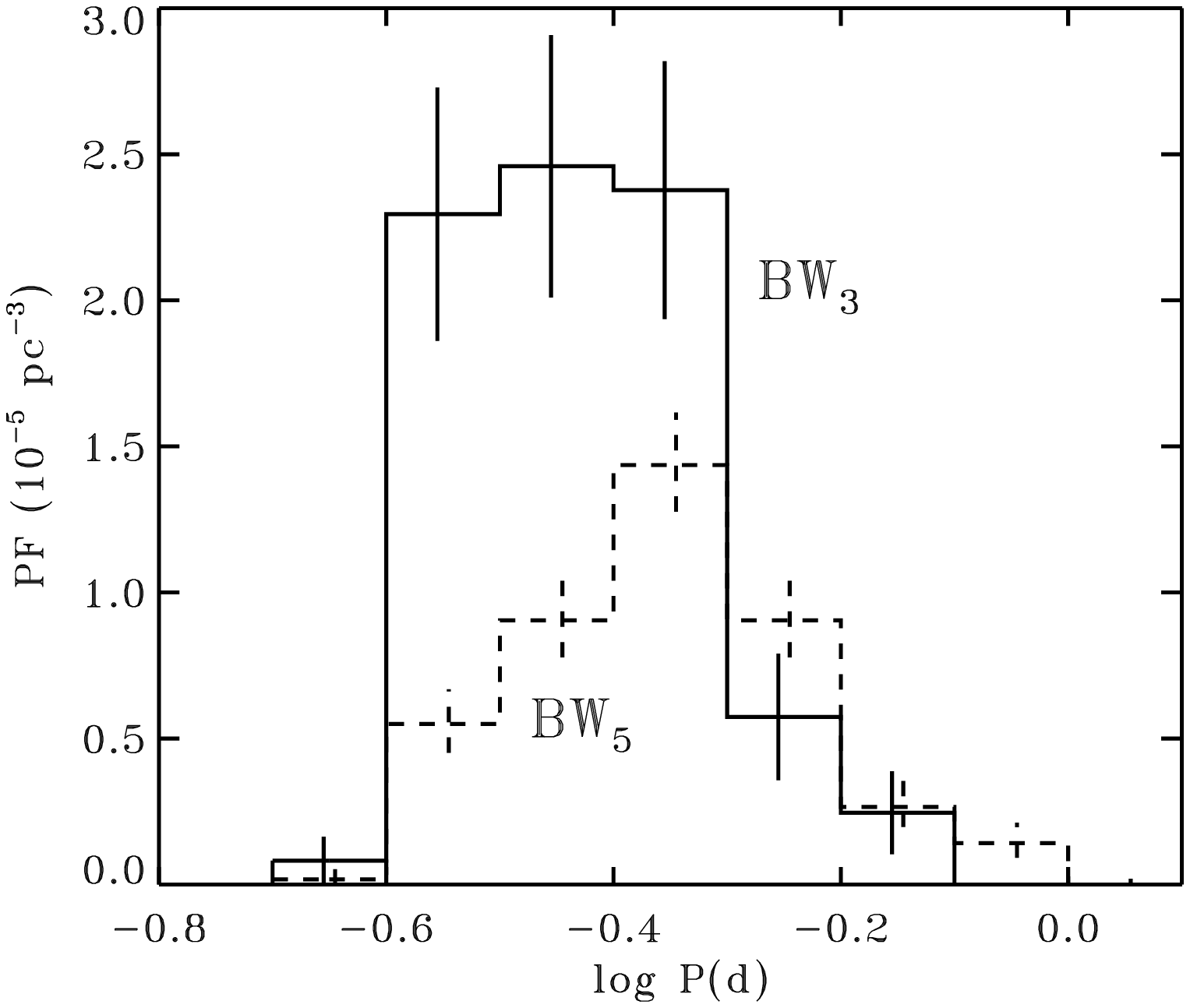] {\label{fig2} 
The ``period function'', PF, giving the number of detected systems in the
BW$_3$ (continuous line) and BW$_5$ (broken line) 
samples in intervals of $\Delta \log P = 0.1$ (in
days) per unit of search volume. The vertical bars give the Poisson
errors.
}

\figcaption[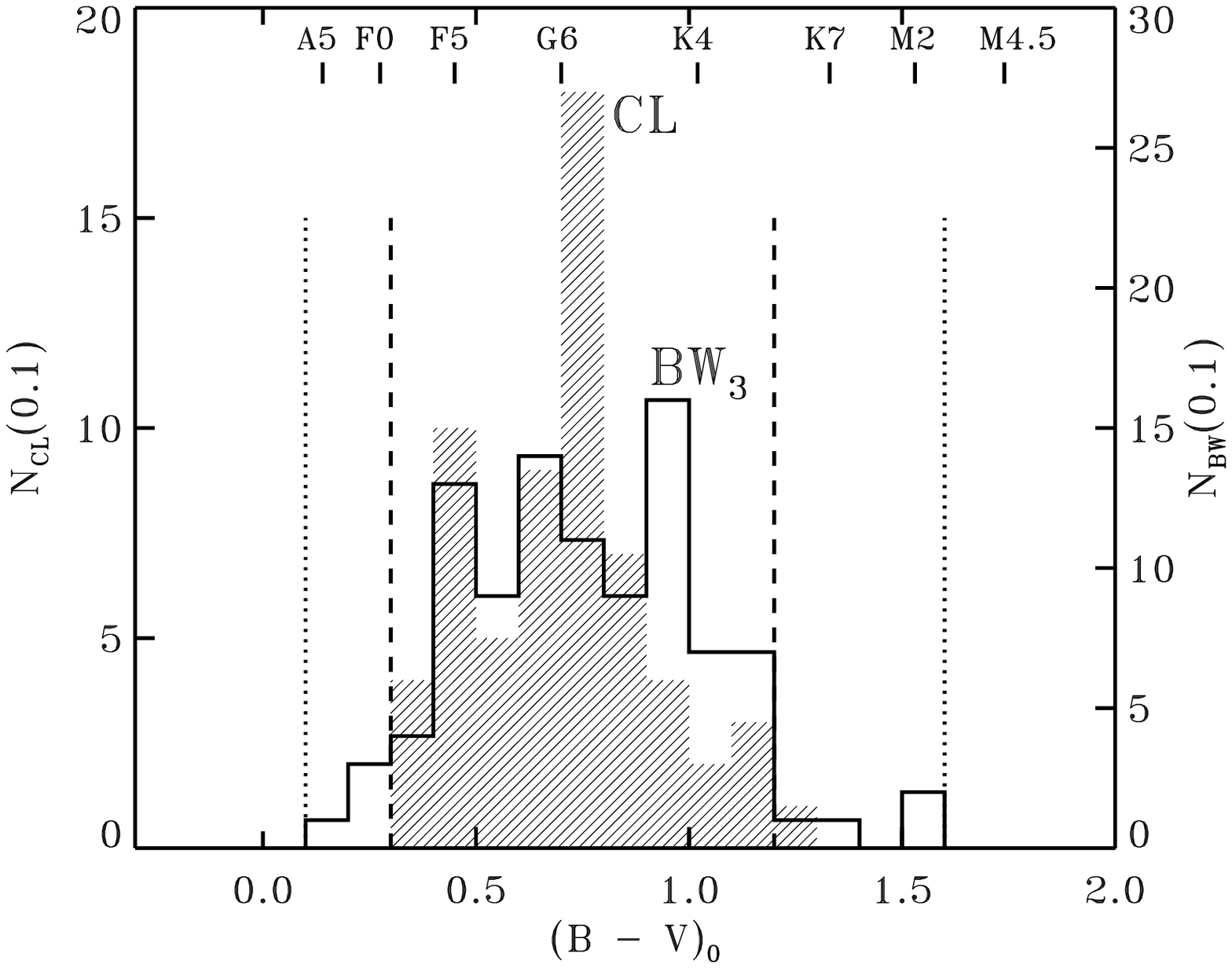] {\label{fig3}
Same as in Figure~\ref{fig1} for
the $(B-V)_0$ color distributions.
The vertical broken and dotted lines define the color ranges
which contain 90 and all 98 systems of the BW$_3$ sample. The
tick marks in the upper edge of the figure give the spectral
types following Bessell (1979).
}
 
\figcaption[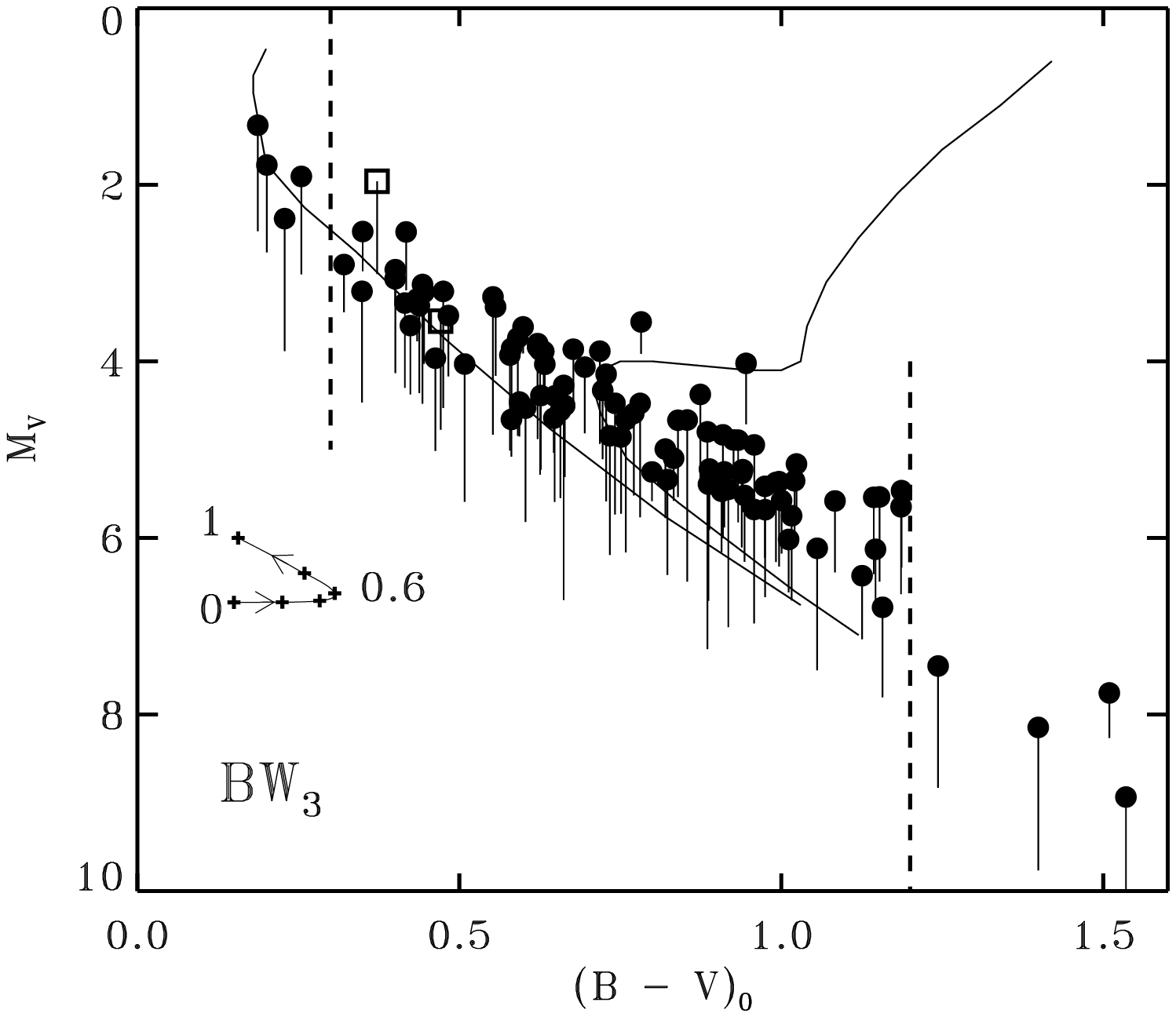] {\label{fig4} 
The color--magnitude diagram for the BW$_3$ sample, with the
observed isochrones for Praesepe and NGC~6791 clusters shown by thin lines. 
The short vectors pointing down from the symbols give the 
light variation amplitudes multiplied by 3 times for 
better visibility. The amplitudes are discussed
in Section~\ref{ampl}. The two systems with unequally
deep eclipses (Poor Thermal Contact binaries) are 
marked by open squares. The vertical broken lines 
mark the color ranges containing 90 percent of the BW$_3$ systems.
The insert at left shows
the expected shifts in both coordinates due to 
the mass-ratio as it changes from $q=0$ to $q=1$;
the tick marks along the line are placed
at intervals of $\Delta q = 0.2$.
}

\figcaption[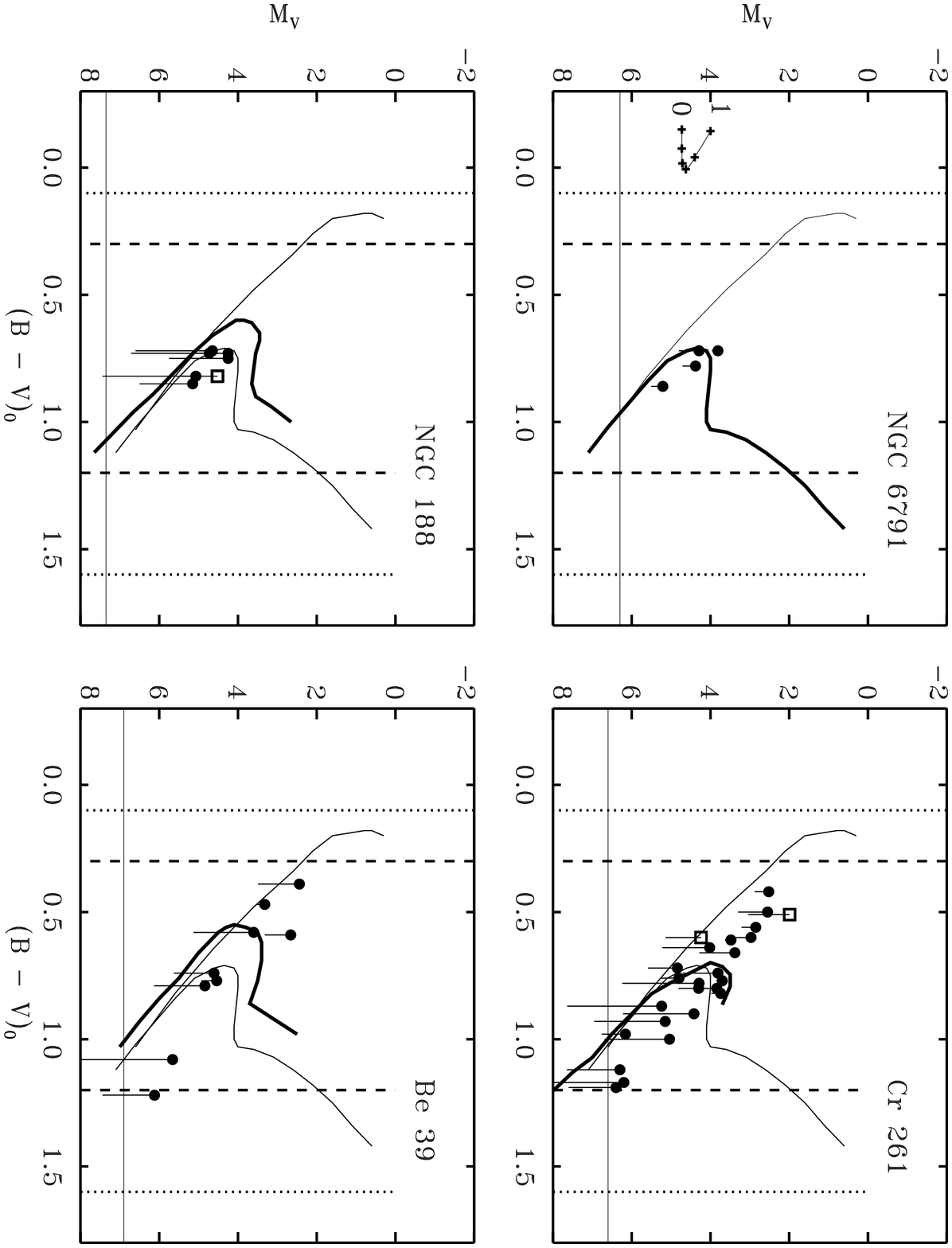] {\label{fig5}
The color--magnitude diagrams for the 4 oldest clusters
which are known to contain contact binaries: NGC~6791, Cr~261,
NGC~188 and Be~39. The vertical broken and dotted lines give
the 90-percent and full ranges for the BW$_3$ sample. The PTC systems
are marked by open squares. Light variation amplitudes
are shown by downward pointing vectors; they
are multiplied by 3 times to improve visibility 
(see Section~\ref{ampl}). The observed isochrones for 
Praesepe and NGC~6791 clusters  are shown by thin lines while the
isochrone of each particular cluster is shown by a thick line.
For each cluster, an estimated limit of the search for 
variability is indicated by a thin horizontal line in the lower
part of the panel.
The first panel gives the expected changes in position due to
the unknown mass ratio, as in Figure~\ref{fig4}.
}

\figcaption[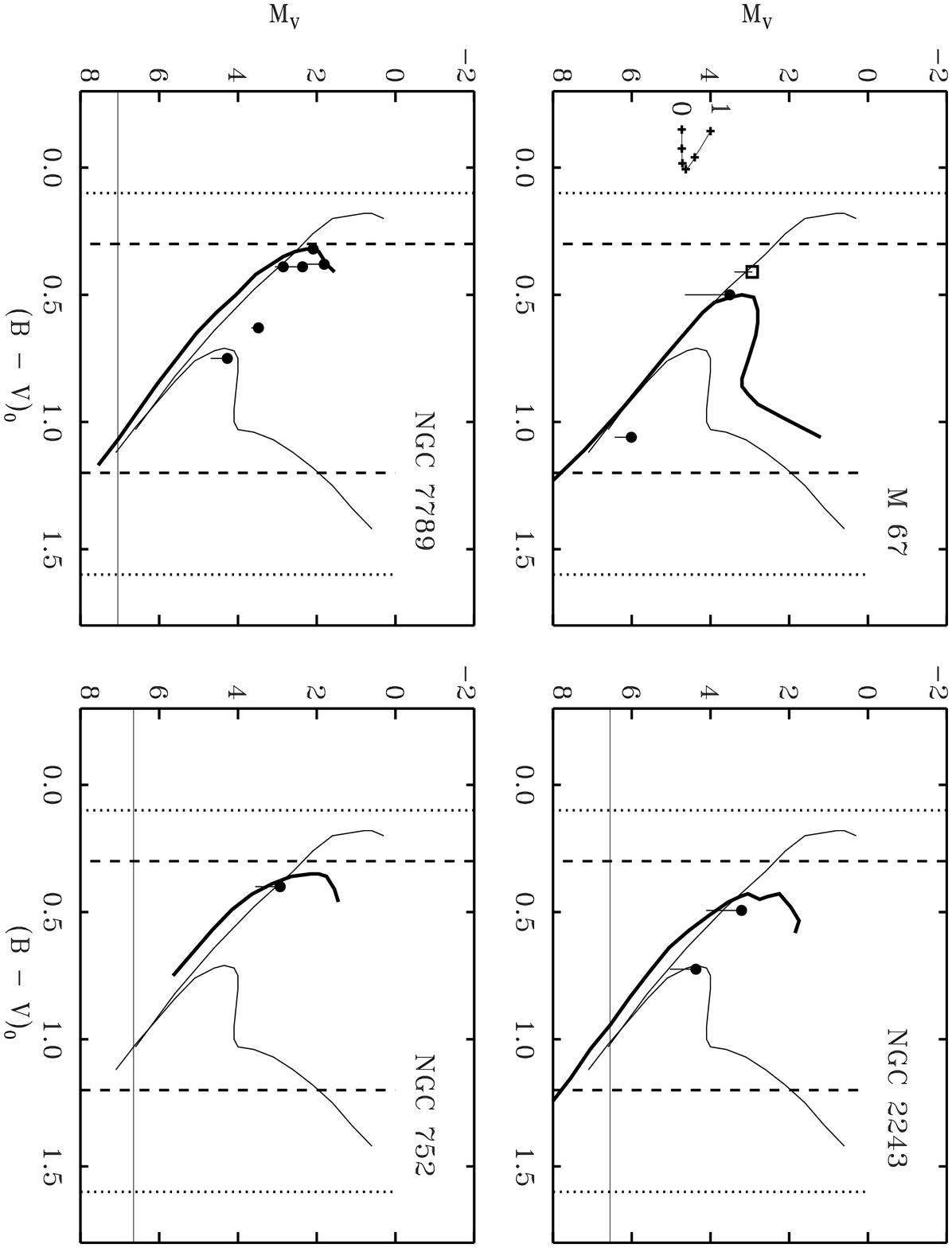] {\label{fig6}
Same as in the previous figure, but for M~67, NGC~2243, NGC~7789 and
NGC~752.
}

\figcaption[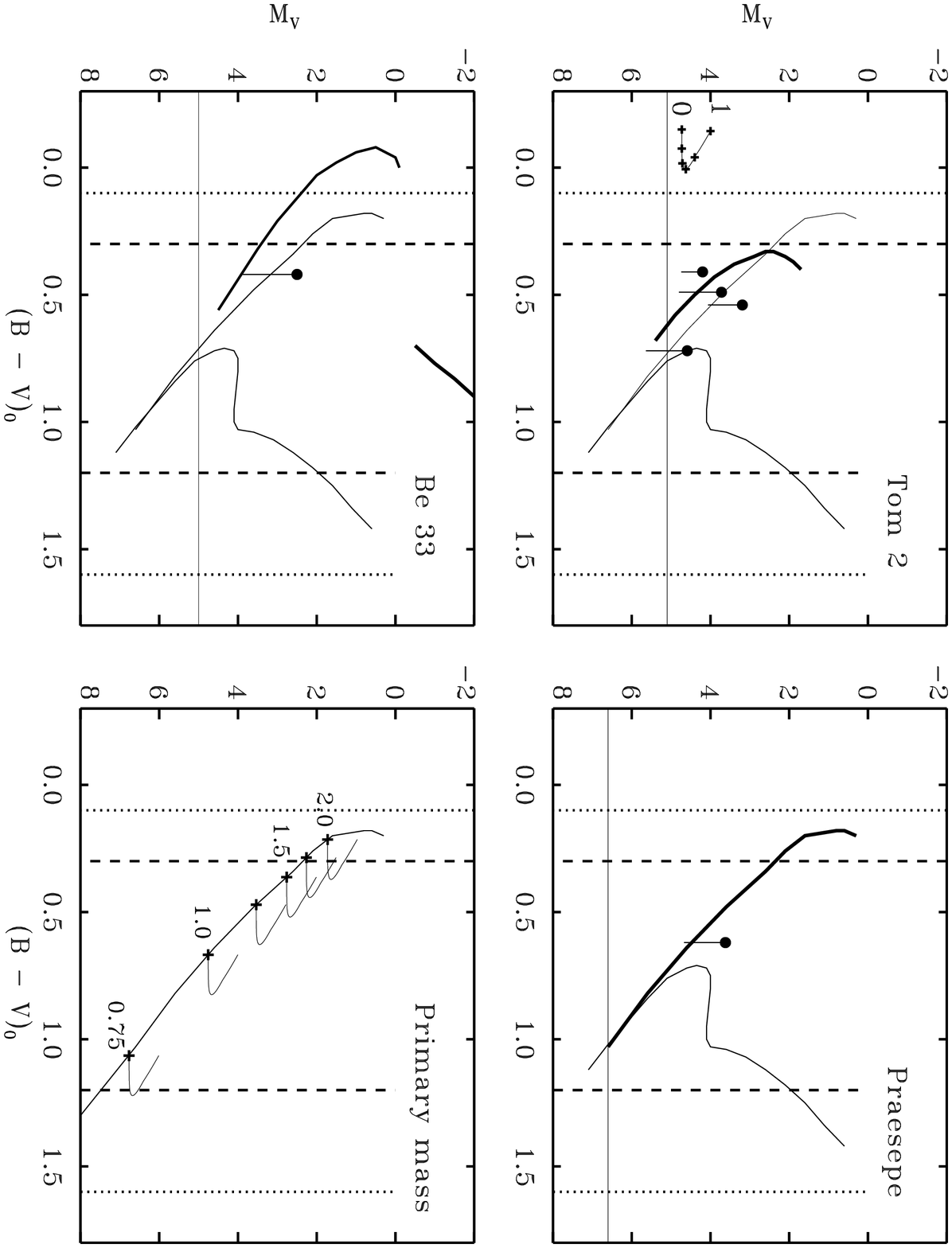] {\label{fig7}
Same as the two previous figures, but for the three ``youngest'' clusters
of the CL sample: Tom~2, Praesepe and Be~33. It is possible that
the only system visible in Be~33 does not actually belong to it. The
last panel of the figure shows the observed isochrone for
Praesepe with the tick marks corresponding to 0.25~M$_\odot$ 
intervals of the mass of the primary component, 
as given by the mass--luminosity calibration for the disk stars
by Kroupa et al.\ (1993). Changes due
to the mass-ratio are shown by short curves which duplicate
the insert in the first panel.
}

\figcaption[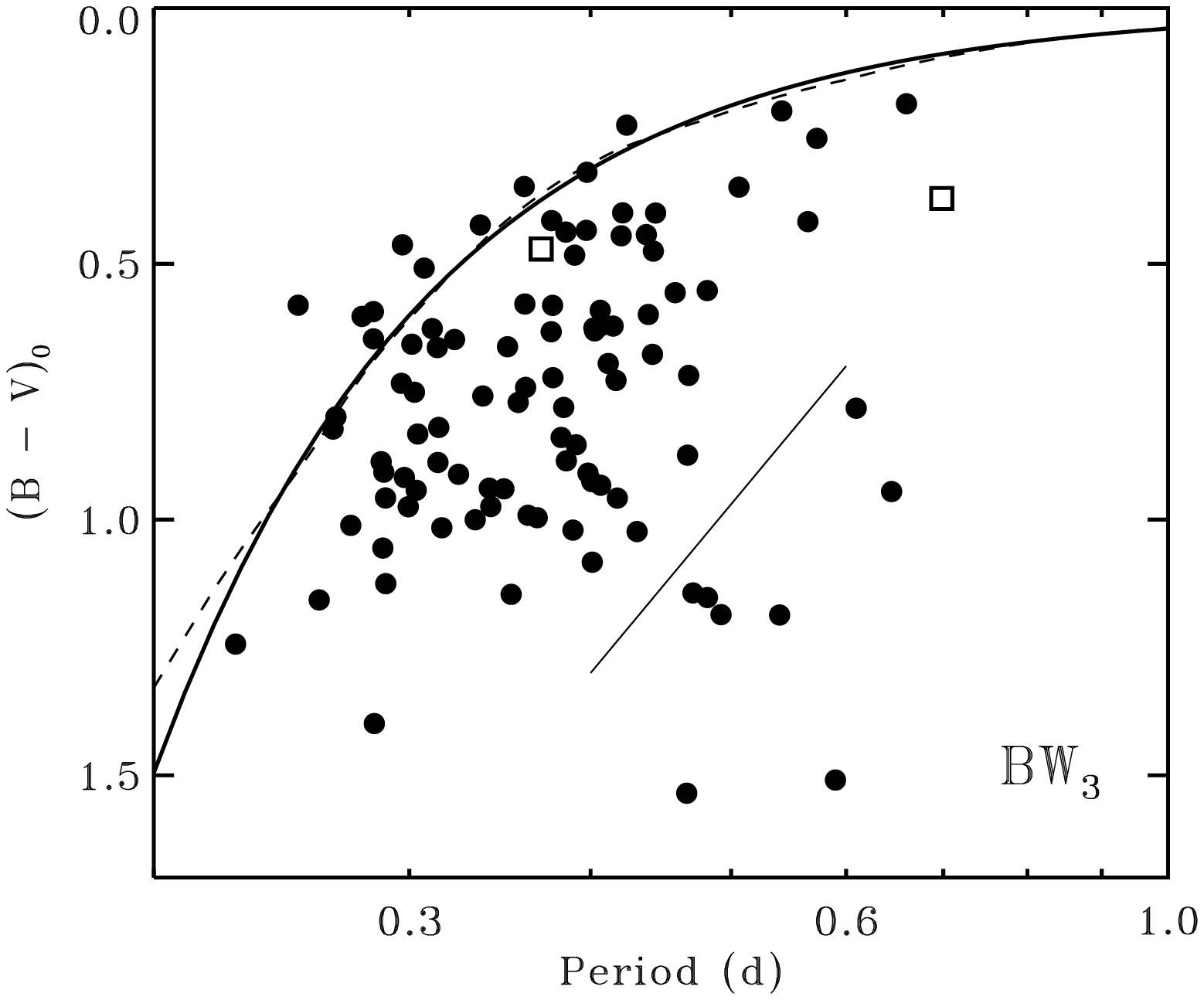] {\label{fig8}
The period -- color relation for the BW$_3$ sample, with data for
contact binaries shown as filled circles. The
open squares mark the two PTC systems in the sample.  
The curve gives the location of the SPBE for the $B-V$ color,
$(B-V)_{SPBE} = 0.04 \times P^{-2.25}$, which was determined
to be consistent with the previously 
introduced in R97a for the $V-I$ color (see the text). 
The previous relation, directly transformed from 
$V-I$ to $B-V$ is shown by a broken line. It is
suspected that some of the systems appearing in the lower right
area, below the slanted line,
may have aliased periods and/or incorrect reddening values.
}

\figcaption[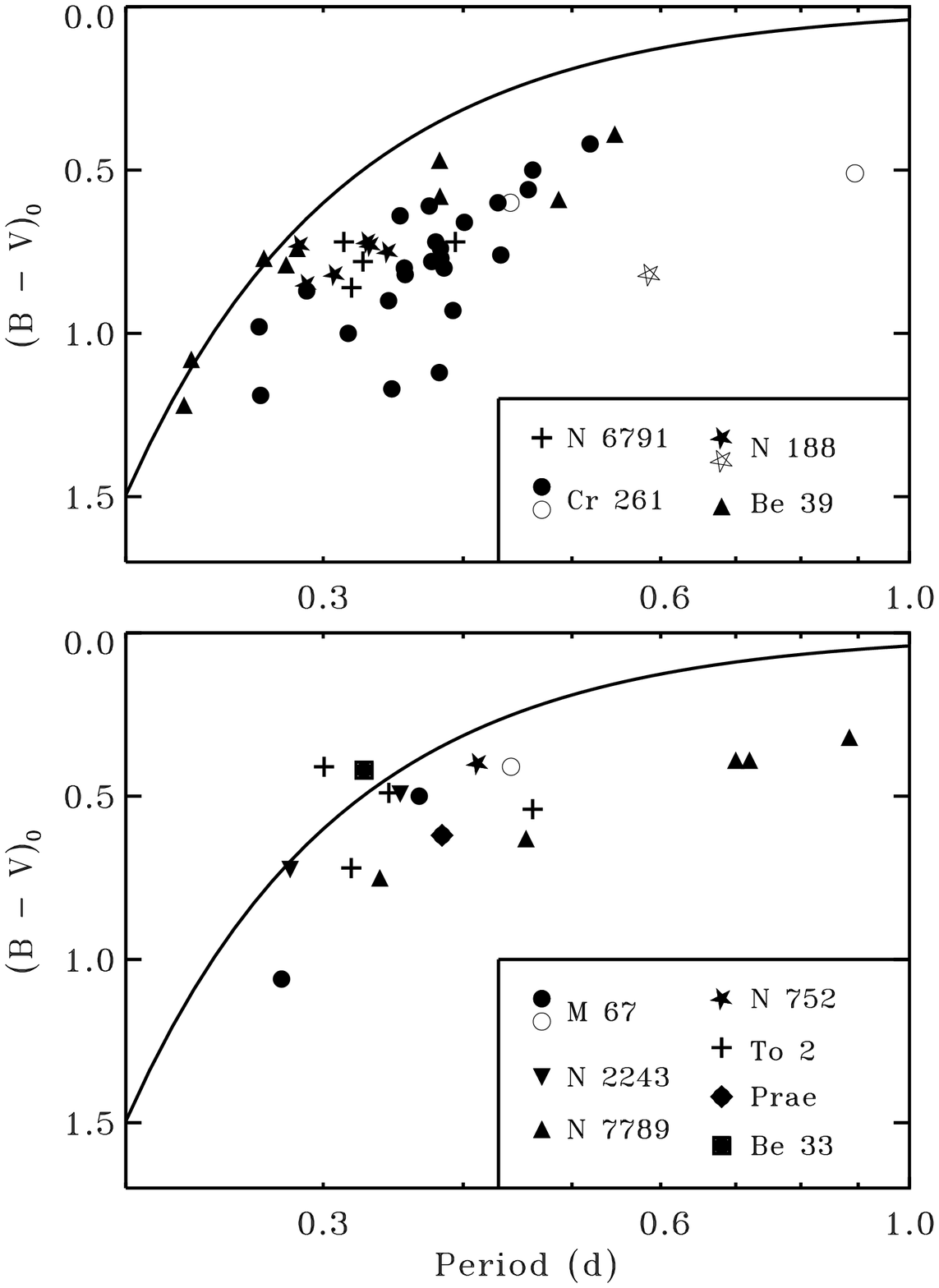] {\label{fig9}
The period -- color relation for the CL sample. 
Different symbols are used for each cluster, as 
explained in the legends to both panels, which give
data separately for the oldest (upper panel) and the ``youngest''
(lower panel) clusters. 
The open symbols mark the PTC systems. The curve gives the
approximation of the SPBE, as in the previous figure.
}

\figcaption[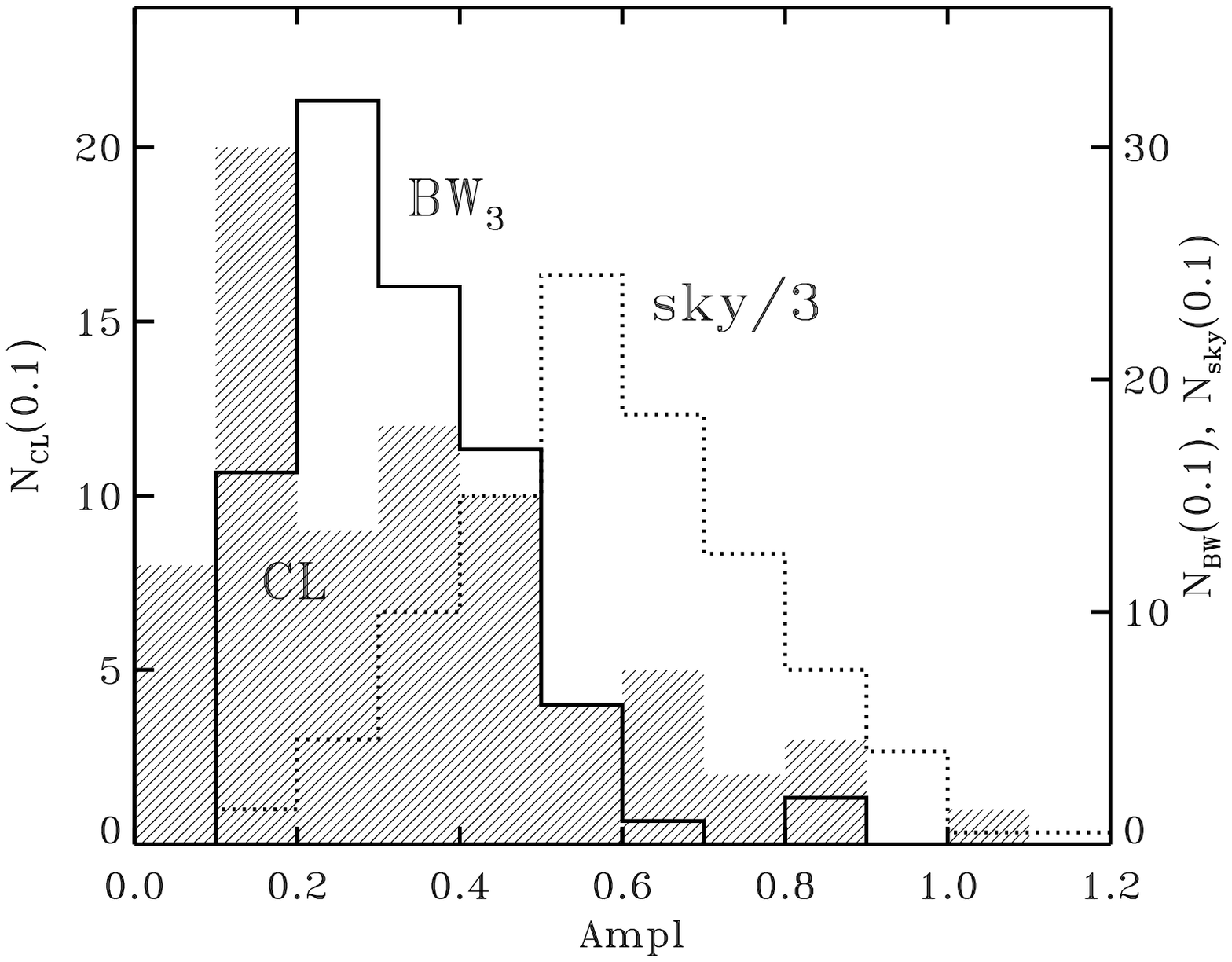] {\label{fig10}
Comparison of the amplitude distributions 
for the BW$_3$ (line), CL (hatched) and sky-field (dotted) samples.
The numbers for the sky sample have been scaled down by 3 times.
Note the different scales used on both sides of the figure.}

\figcaption[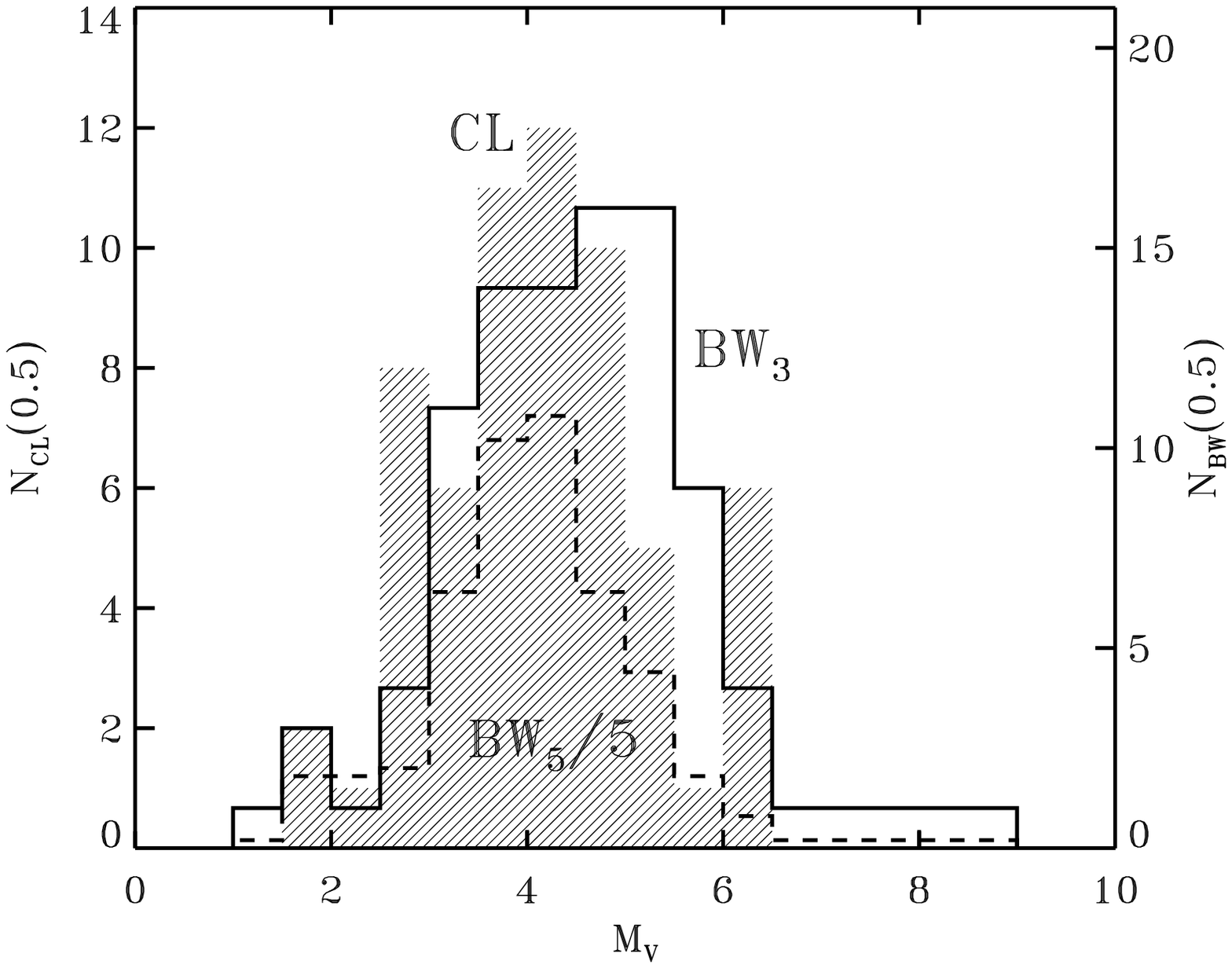] {\label{fig11}
The distributions of the absolute magnitudes in the BW and CL samples.
The format is similar to that used in Figures~\ref{fig1} and \ref{fig2}.
The broken line gives the shape of the
distribution for the BW$_5$ sample after scaling down by 5 times. 
The luminosity functions derived from both BW distributions
are shown in the next figure.
}

\figcaption[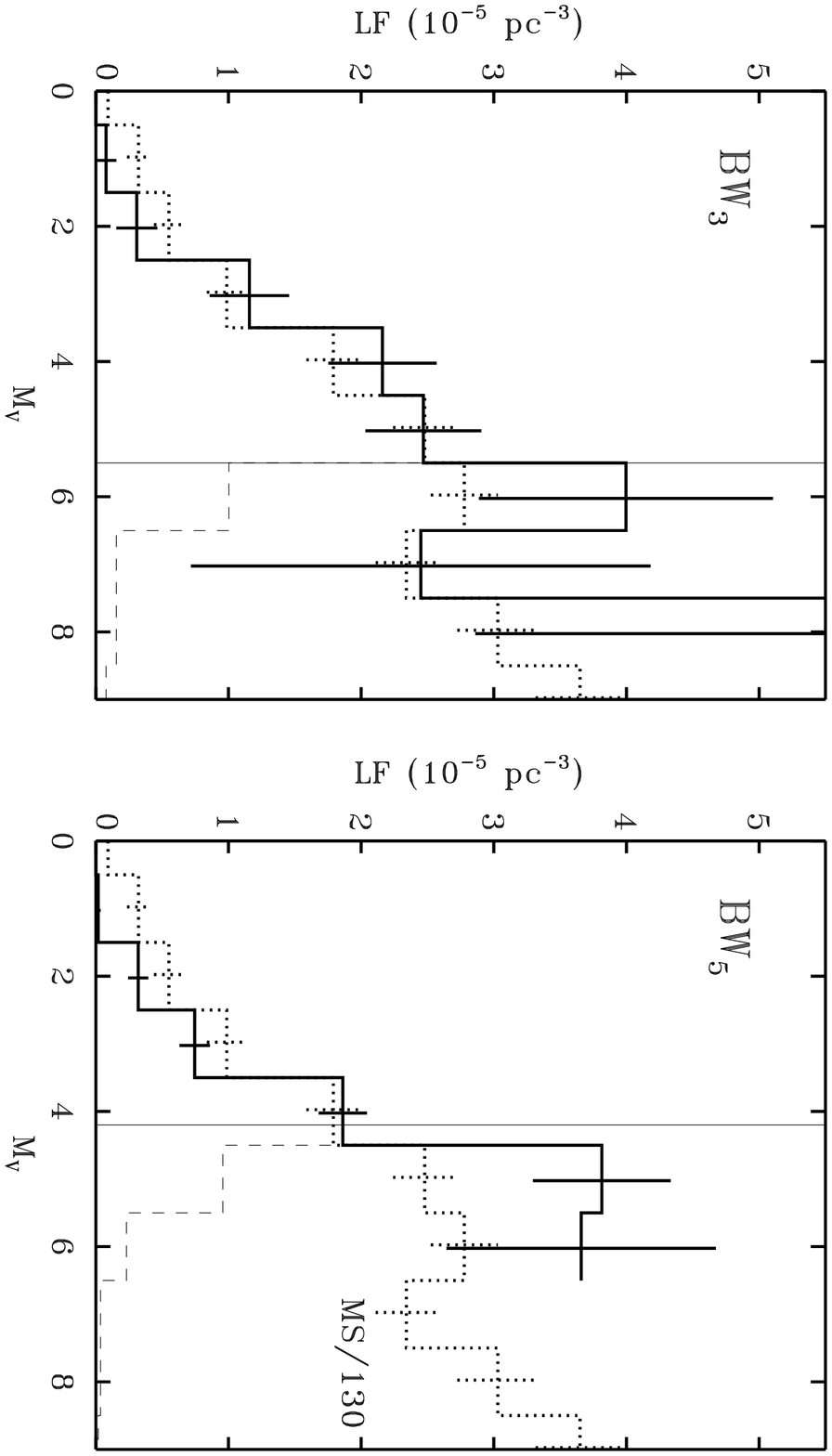] {\label{fig12}
The directly observed luminosity functions derived from the basic
3 kpc BW$_3$ sample (left panel) and the deeper
5 kpc BW$_5$ sample (right panel) are shown by continuous lines. 
The luminosity function for the local main sequence stars 
LF$_{MS}$, scaled down by a factor of 130, is plotted
for comparison (dotted line in both panels).
The vertical bars give the Poisson errors for individual
bins. Beyond $M_V > 5.5$ for BW$_3$ and $M_V > 4.5$ for
BW$_5$, the plotted data include corrections for
the progressively decreasing volumes; 
the uncorrected data are shown by thin broken lines. The bin
$3.5 < M_V < 4.5$ of BW$_5$ is partly affected by the finite depth
of the search which is complete to $M_V \simeq 4.2$; this difference
has been disregarded here. Note that the luminosity functions
shown here are not been corrected for the structure of the galactic
disk. Such corrections require an assumption on the
population characteristics of the contact binaries. 
}

\figcaption[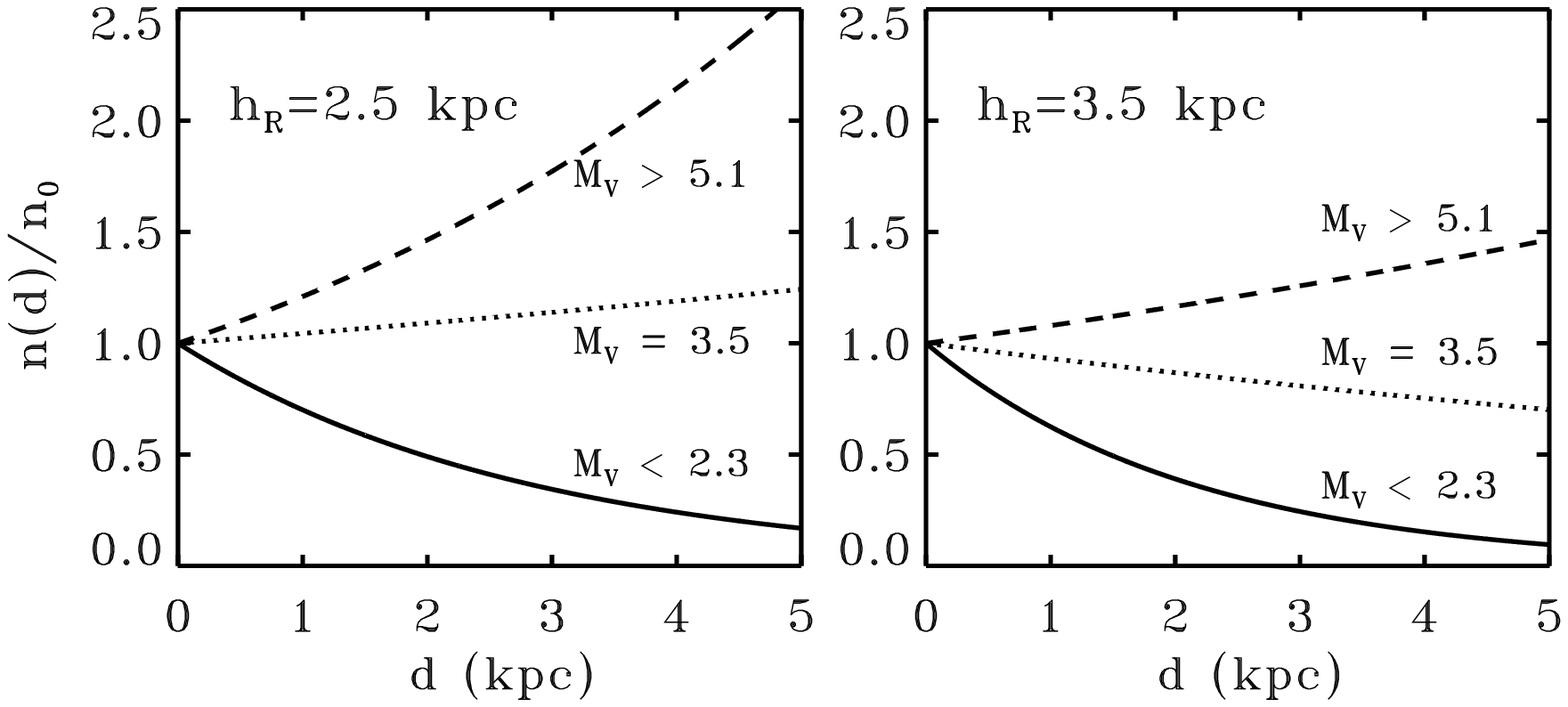] {\label{fig13}
Ratios of the MS star number density for the galactic disk models
to the local MS star density, for two values of the 
length scale $h_R$ = 2.5 and 3.5 kpc (left and right panels)
and for three representative values of the scale height $h_z (M_V)$ =
90, 190 and 325 pc.
}

\figcaption[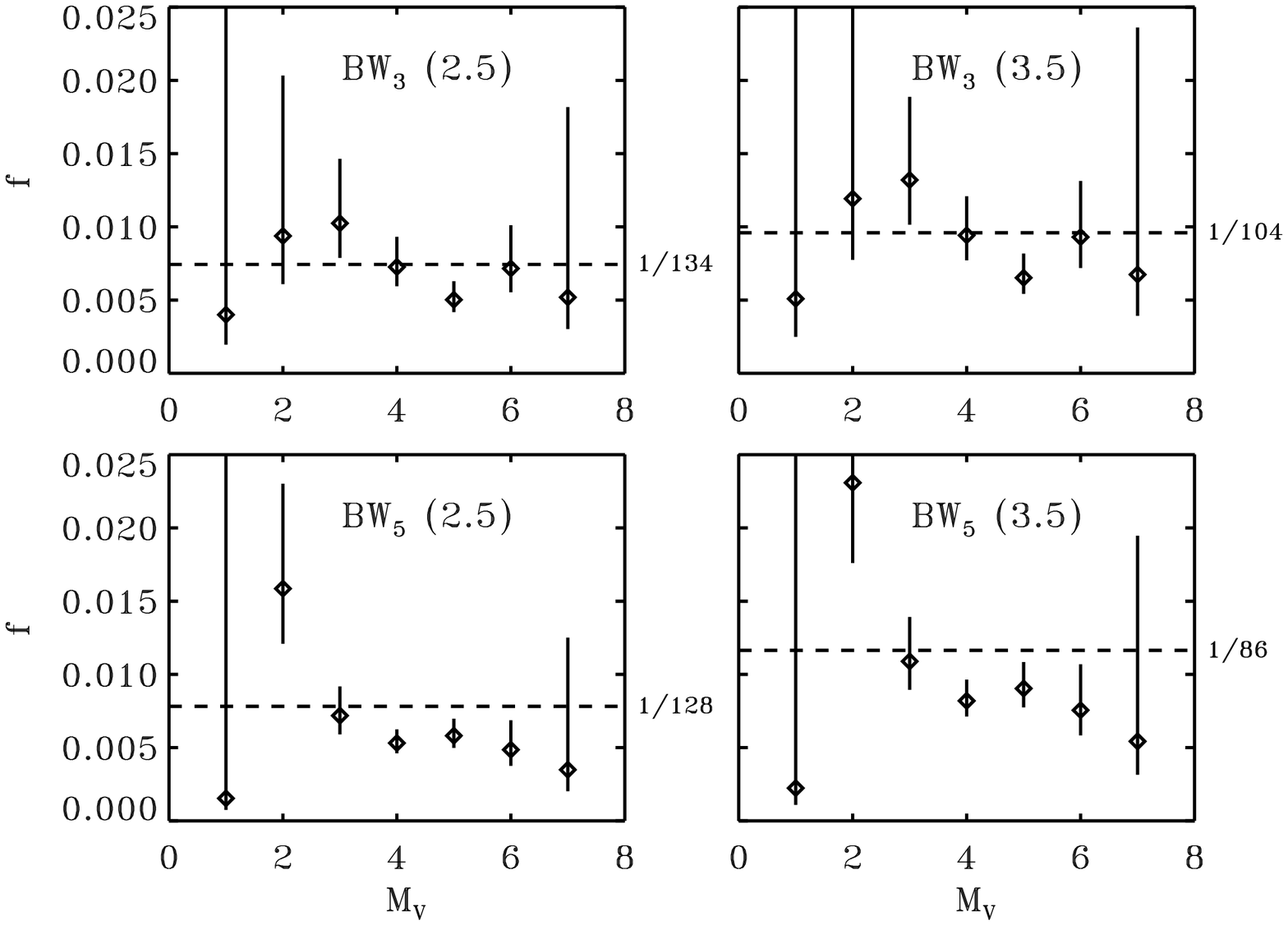] {\label{fig14}
Apparent frequency of contact binary systems, $f$, relative to the 
number of MS stars
determined from the galactic structure corrected
luminosity functions LF$_{BW}^{corr}$, for the two BW samples
(3 kpc, upper row of panels; 5 kpc, lower row of panels) 
and for two values
of the galactic length scale $h_R$ (2.5 kpc, left panels; 3.5 kpc,
right panels). The weighted mean values 
of the frequencies are shown by broken lines.
The vertical bars give the errors determined for the
inverse frequencies, as listed in Table~7.
}

\figcaption[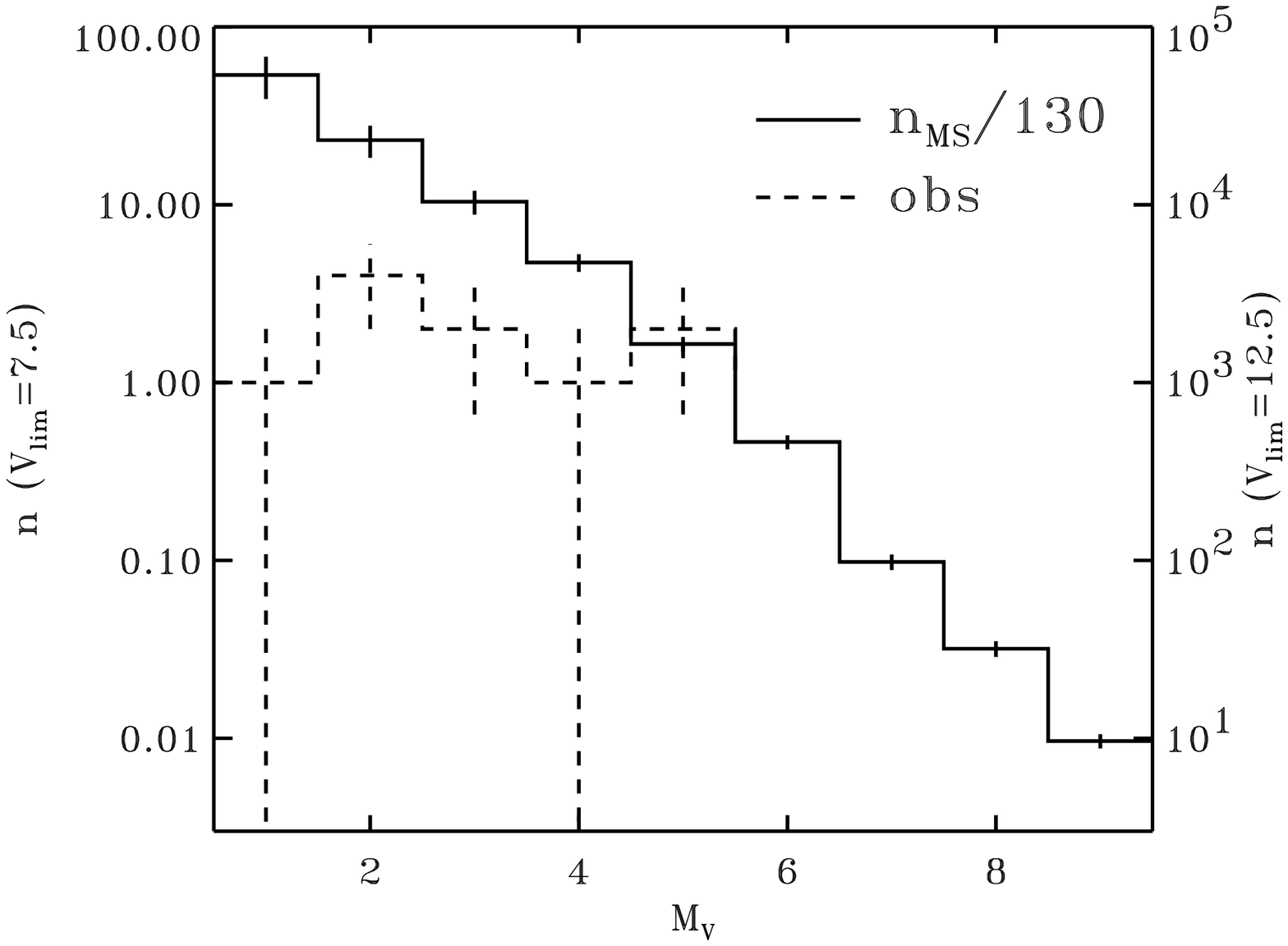] {\label{fig15}
The expected numbers of contact systems in one magnitude
wide bins of $M_V$, in the whole sky to the limiting 
magnitudes $V_{lim} = 7.5$ (left vertical axis) and 
$V_{lim} = 12.5$ (right vertical axis), 
predicted from the main-sequence
luminosity function and scaled down by 130, are
shown by the continuous line histogram. The currently known
numbers of the systems to $V_{lim} = 7.5$
are shown by a broken line. The non-detections are plotted
at the level of  $1 \pm 1$ which is 
plausible from the point of view of
the Poissonian-statistics in the adjacent bins. The figure suggests
a decrease in the frequency of occurrence for $M_V < 3.5$,
whereas the BW data (see Figure~14) are consistent with the frequency
staying constant for $M_V > 2.5$.
}

\figcaption[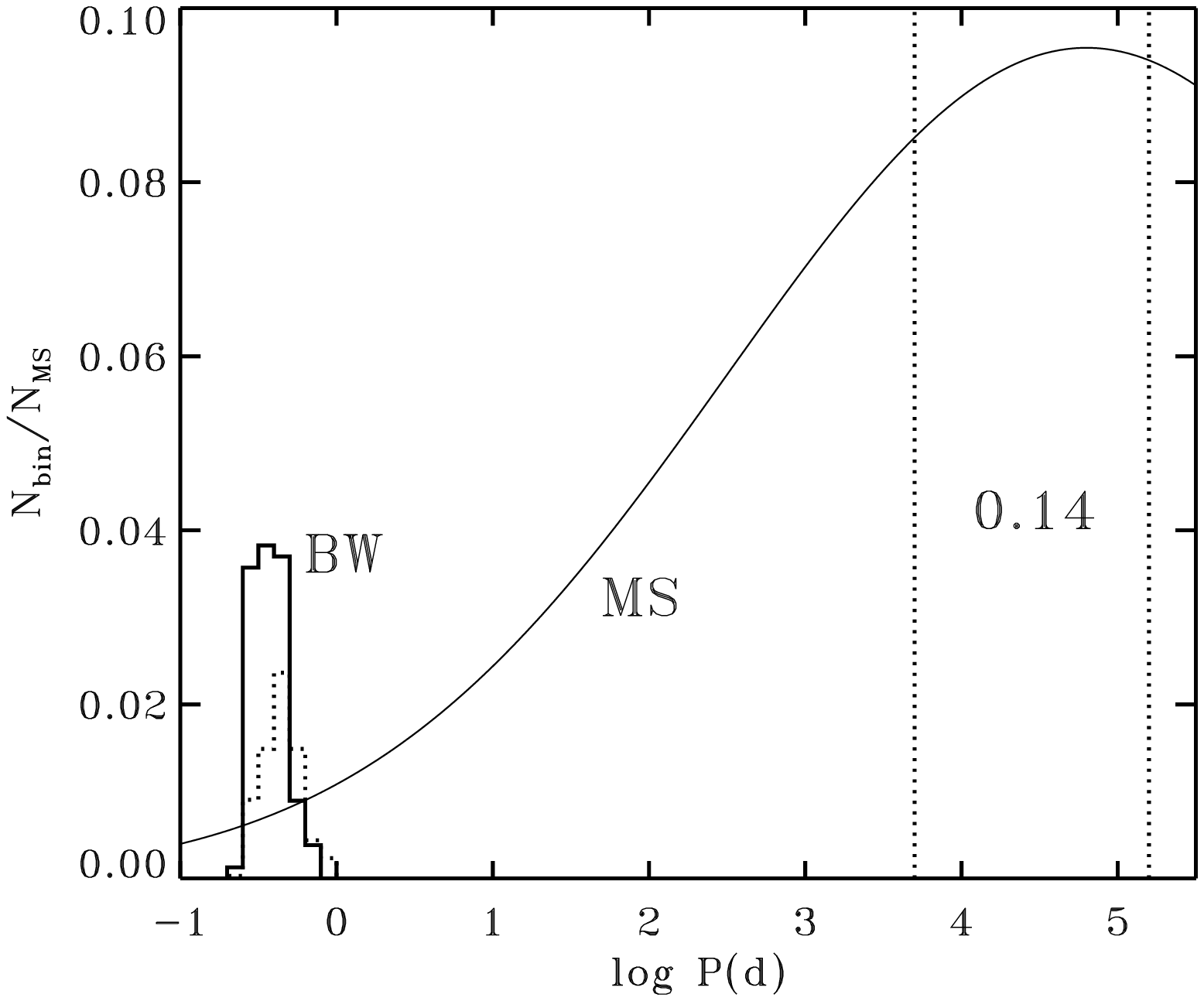] {\label{fig16}
The observed spatial (free of geometric and inclination 
effects) frequency of occurrence of disk binary systems 
following the study of Duquennoy \& Mayor (1991) 
is shown as one branch of a Gaussian curve, with
the logarithmic units of the period in days. It has 
been normalized to give the integrated 
observed frequency in the range $3.7 < \log P(d) < 5.2$ 
of 0.14, following the study of Patience et al.\ (1998). 
The histograms for the contact systems in the
BW$_3$ (continuous line ) and BW$_5$ (broken line) 
samples have been copied from the period function in 
Figure~\ref{fig2} after normalization to give 
the integrated spatial frequency of occurrence of contact
binaries equal to 1/80 (about 1.5 higher than the apparent frequency).
This conversion is approximate with the combined 
uncertainty in the spatial frequency of contact binaries
relative to main sequence stars of about 50 percent. 
} 

\begin{table}
\dummytable\label{tab1}
\end{table}

\begin{table}
\dummytable\label{tab2}
\end{table}

\begin{table}
\dummytable\label{tab3}
\end{table}

\begin{table}
\dummytable\label{tab4}
\end{table}

\begin{table}
\dummytable\label{tab5}
\end{table}

\begin{table}
\dummytable\label{tab6}
\end{table}

\begin{table}
\dummytable\label{tab7}
\end{table}

\end{document}

\begin{deluxetable}{llccccccc}           % Table 3
\small
\tablecolumns{8}
%\tablewidth{360pt}
\tablewidth{0pt}
\tablenum{3}
\tablecaption{W~UMa-type Systems in Open Clusters (CL)} 
\tablehead{
  \colhead{Name$_1$} & \colhead{Name$_2$} &
  \colhead{P(day)}   & \colhead{$V_{max}$} & 
  \colhead{$B-V$}    & \colhead{$M_V$}    &
  \colhead{$\Delta M_V$} & 
  \colhead{Ampl$_V$} & \colhead{Comm}  
}
\startdata
\sidehead{\bf NGC 6791} 
V513 Lyr & V7 & 0.3935 & 17.69 & 0.87 & 4.19 & \phs 0.16 & 0.17 &    \nl
V514 Lyr & V8 & 0.4000 & 17.66 & 1.27 & 4.16 & $-$1.05   & 0.11 & nm \nl
V518 Lyr & V5 & 0.3130 & 17.21 & 0.87 & 3.71 & $-$0.76   & 0.05 &    \nl
V519 Lyr & V1 & 0.2677 & 16.22 & 0.85 & 2.72 & $-$1.99   & 0.40 & nm \nl
V521 Lyr & V4 & 0.3255 & 17.78 & 0.93 & 4.28 & $-$0.30   & 0.11 &    \nl
V522 Lyr & V6 & 0.2797 & 15.45 & 0.70 & 1.95 & $-$2.23   & 0.12 & nm \nl
V525 Lyr & V3 & 0.3180 & 18.61 & 1.01 & 5.11 & \phs 0.24 & 0.10 &    \nl
V526 Lyr & V2 & 0.3170 & 19.70 & 0.98 & 6.20 & \phs 1.42 & 0.10 & nm \nl
\sidehead{\bf Cr~261}
GV Mus & V36 & 0.4320 &  17.80 & 0.98 & 4.80 & \phs 0.77 &   0.15 &    \nl
GW Mus & V41 & 0.2630 &  19.16 & 1.20 & 6.16 & \phs 0.51 &   0.20 &    \nl
GX Mus &  V4 & 0.3817 &  16.81 & 0.96 & 3.81 & $-$0.40   &   0.30 &    \nl
GY Mus &  V2 & 0.3550 &  16.74 & 1.04 & 3.74 & $-$0.85   &   0.08 &    \nl
GZ Mus & V30 & 0.3513 &  17.02 & 0.86 & 4.02 & $-$0.05   &   0.40 &    \nl
HK Mus & V10 & 0.3808 &  19.30 & 1.34 & 6.30 & \phs 0.94 &   0.45 &    \nl
HL Mus & V26 & 0.4613 &  15.55 & 0.72 & 2.55 & $-$0.57   &   0.25 &    \nl
HM Mus & V19 & 0.3431 &  17.42 & 1.12 & 4.42 & $-$0.48   &   0.60 &    \nl
HO Mus & V17 & 0.4571 &  15.85 & 0.78 & 2.85 & $-$0.47   &   0.12 &    \nl
HQ Mus & V16 & 0.8841 &  14.99 & 0.73 & 1.99 & \phs 0.09 &   0.35 &    \nl
HU Mus & V32 & 0.5190 &  15.52 & 0.64 & 2.52 & $-$0.13   &   0.12 &    \nl
HV Mus & V31 & 0.3820 &  16.70 & 0.99 & 3.70 & $-$0.60   &   0.05 &    \nl
HX Mus & V35 & 0.3697 &  19.42 & 1.12 & 6.42 & \phs 1.66 &   0.60 & nm \nl
HY Mus &  V9 & 0.4114 &  18.67 & 1.06 & 5.67 & \phs 1.30 &   0.25 & nm \nl
HZ Mus & V37 & 0.4295 &  15.97 & 0.82 & 2.97 & $-$0.59   &   0.13 &    \nl
II Mus & V28 & 0.3845 &  17.30 & 1.02 & 4.30 & $-$0.08   &   0.17 &    \nl
IK Mus & V13 & 0.3749 &  17.29 & 1.00 & 4.29 & $-$0.08   &   0.65 &    \nl
IL Mus & V33 & 0.2900 &  18.24 & 1.09 & 5.24 & \phs 0.11 &   0.80 &    \nl
IO Mus & V25 & 0.4009 &  16.38 & 0.88 & 3.38 & $-$0.50   &   0.30 &    \nl
IR Mus & V15 & 0.3158 &  18.04 & 1.22 & 5.04 & $-$0.32   &   0.55 &    \nl
IS Mus & V18 & 0.3454 &  19.20 & 1.39 & 6.20 & \phs 0.50 &   0.80 &    \nl
IT Mus & V34 & 0.3727 &  18.90 & 0.99 & 5.90 & \phs 1.55 &   0.20 & nm \nl
IU Mus & V14 & 0.2638 &  19.40 & 1.41 & 6.40 & \phs 0.12 &   0.40 &    \nl
IW Mus & V24 & 0.3544 &  16.84 & 1.02 & 3.84 & $-$0.70   &   0.13 &    \nl
IX Mus &  V1 & 0.3780 &  17.84 & 0.94 & 4.84 & \phs 0.67 &   0.25 &    \nl
IZ Mus &  V3 & 0.3916 &  18.15 & 1.15 & 5.15 & \phs 0.41 &   0.60 &    \nl
KL Mus & V43 & 0.3730 &  16.48 & 0.83 & 3.48 & $-$0.38   &   0.05 &    \nl
KM Mus & V39 & 0.4307 &  17.24 & 0.82 & 4.24 & \phs 0.68 &   0.30 &    \nl
\sidehead{\bf NGC~188}
EP Cep &      & 0.2897 & 16.50 & 0.95 & 5.00 & $-$0.08   &   0.45 &    \nl
EQ Cep &      & 0.3069 & 16.42 & 0.92 & 4.92 & \phs 0.05 &   0.79 &    \nl
ER Cep &      & 0.2857 & 15.60 & 0.83 & 4.10 & $-$0.64   &   0.82 &    \nl
ES Cep &      & 0.3424 & 15.60 & 0.85 & 4.10 & $-$0.35   &   0.50 &    \nl
V369 Cep & V7 & 0.3282 & 16.00 & 0.82 & 4.50 & \phs 0.06 &   0.65 &    \nl
V370 Cep & V6 & 0.3304 & 16.08 & 0.83 & 4.58 & \phs 0.12 &   0.15 &    \nl
V371 Cep & V5 & 0.5860 & 15.87 & 0.92 & 4.37 & \phs 0.74 &   0.33 &  
\tablebreak

\sidehead{\bf Be~39}
V701 Mon & V10 & 0.5460 & 15.94 & 0.50 & \phs 2.54 & \phs 0.11 & 0.35 &    \nl
V702 Mon & V11 & 0.2254 & 19.62 & 1.33 & \phs 6.22 & $-$0.43   & 0.44 &    \nl
V703 Mon &  V8 & 0.2288 & 19.16 & 1.19 & \phs 5.76 & $-$0.44   & 0.77 &    \nl
V704 Mon &  V7 & 0.2780 & 18.34 & 0.90 & \phs 4.94 & \phs 0.00 & 0.43 &    \nl
V705 Mon &  V3 & 0.3810 & 16.82 & 0.58 & \phs 3.42 & \phs 0.05 & 0.07 &    \nl
V706 Mon &  V2 & 0.4865 & 16.16 & 0.70 & \phs 2.76 & $-$0.50   & 0.22 &    \nl
V707 Mon &  V9 & 0.2598 & 20.12 & 0.95 & \phs 6.72 & \phs 1.49 & 1.00 & nm \nl
V709 Mon &  V4 & 0.3813 & 17.10 & 0.69 & \phs 3.70 & \phs 0.00 & 0.51 &    \nl
V710 Mon &  V1 & 0.4052 & 15.36 & 0.58 & \phs 1.96 & $-$1.29   & 0.41 & nm \nl
V711 Mon &  V5 & 0.2656 & 18.04 & 0.88 & \phs 4.64 & $-$0.33   & 0.13 &    \nl
V712 Mon &  V6 & 0.2844 & 18.11 & 0.85 & \phs 4.71 & $-$0.04   & 0.34 &    \nl
\sidehead{\bf M~67}
AH Cnc & III$-$33 & 0.3604 & 13.31 & 0.54 & \phs 3.72 & \phs 0.15 & 0.38 & \nl
ET Cnc & III$-$79 & 0.2704 & 15.81 & 1.10 & \phs 6.22 & \phs 0.41 & 0.14 & \nl
EV Cnc &  III$-$2 & 0.4412 & 12.74 & 0.45 & \phs 3.15 & \phs 0.24 & 0.15 & \nl
\sidehead{\bf NGC~2243}
         & V2 & 0.2853 &  17.82 & 0.78 & \phs 4.72 & \phs 0.02 & 0.22 & nd \nl
         & V3 & 0.3564 &  16.66 & 0.55 & \phs 3.56 & $-$0.01   & 0.30 & nd \nl
\sidehead{\bf NGC~7789}
V730 Cas & V6 & 0.8840 &  14.54 & 0.62 & \phs 2.24 & \phs 0.73 & 0.05 &    \nl
V732 Cas & V1 & 1.1900 &  14.26 & 0.68 & \phs 1.96 & \phs 0.85 & 0.16 &    \nl
V733 Cas & V7 & 0.4550 &  15.93 & 0.93 & \phs 3.63 & $-$0.09   & 0.06 &    \nl
V736 Cas & V2 & 0.7200 &  14.81 & 0.69 & \phs 2.51 & \phs 0.40 & 0.16 &    \nl
V737 Cas & V4 & 0.3370 &  16.72 & 1.05 & \phs 4.42 & $-$0.24   & 0.14 &    \nl
V738 Cas & V5 & 0.2387 &  19.32 & 1.20 & \phs 7.02 & \phs 1.24 & 0.40 & nm \nl
         & V3 & 0.7000 &  15.30 & 0.69 & \phs 3.00 & \phs 0.83 & 0.07 & nd \nl
\sidehead{\bf NGC~752}
QX And & H235 & 0.4118 &  11.28 & 0.44 & \phs 2.93 & $-$0.14   & 0.21 &    \nl
\sidehead{\bf Tom~2}
         & V2 & 0.3278 &  20.19 & 1.02 & \phs 4.69 & \phs 0.55 & 0.35 & nd \nl
         & V3 & 0.4712 &  18.79 & 0.84 & \phs 3.29 & \phs 0.39 & 0.29 & nd \nl
         & V4 & 0.3105 &  19.80 & 0.71 & \phs 4.30 & \phs 0.99 & 0.18 & nd \nl
         & V5 & 0.3533 &  19.32 & 0.79 & \phs 3.82 & \phs 0.52 & 0.36 & nd \nl
\sidehead{\bf Praesepe}
TX Cnc   &    & 0.3829 &  10.02 & 0.62 & \phs 3.78 & $-$0.06 & 0.35 &    \nl
\sidehead{\bf Be~33}
II CMa   & V1 & 0.2292 &  15.22 & 1.30 & $-$0.28   & $-$5.05 & 0.50 & nm \nl
IK CMa   & V2 & 0.3262 &  18.00 & 1.12 & \phs 2.50 & $-$1.05 & 0.46 & xx \nl
\tablecomments{Last column:
nm -- non member; nd -- no designation as variable star available;
xx -- formally not fulfilling the membership criterion,
of $|\Delta M_V| < 1$ mag, but included in the CL sample.
$\Delta M_V = M_V - M_V^{cal}$, where $M_V = V_{max}-(m-M)_V$,
the distance moduli are given in Table~2 and 
$M_V^{cal}$ come from the Hipparcos calibration.}
\enddata
\end{deluxetable}